\documentclass[fleqn,usenatbib]{mnras}

\usepackage[T1]{fontenc}

\DeclareRobustCommand{\VAN}[3]{#2}
\let\VANthebibliography\thebibliography
\def\thebibliography{\DeclareRobustCommand{\VAN}[3]{##3}\VANthebibliography}

\usepackage{graphicx}	
\usepackage{amsmath}	
\usepackage{amssymb}	
\usepackage{newtxtext,newtxmath}
\newcommand{\be}{\begin{equation}}
\newcommand{\ee}{\end{equation}}
\newcommand{\beal}{\begin{aligned}}
\newcommand{\eeal}{\end{aligned}}
\newcommand{\Msol}{\mathrm{M_\odot}}
\usepackage{esdiff}
\usepackage{array}
\usepackage{gensymb}
\usepackage{booktabs}
\usepackage{xcolor}
\usepackage{soul}
\usepackage{ulem}
\usepackage{ragged2e}

\newcommand{\PV}{}
\newcommand{\FC}{}

\newcommand{\refe}{}
\newcommand{\refee}{}

\usepackage{orcidlink}

\title[\refe{Impact of a BBH on its outer circumbinary disc}]
{\refe{Impact of a binary black hole on its outer circumbinary disc}}

\author[R. Mignon-Risse et al.]{
Raphaël Mignon-Risse\,\orcidlink{0000-0002-3072-1496},$^{1}$\thanks{E-mail: raphael.mignon-risse@apc.in2p3.fr}
Peggy Varniere\,\orcidlink{0000-0001-8888-5971},$^{2,3}$
Fabien Casse\,\orcidlink{0000-0002-8156-7628}$^{2}$
\\
$^{1}$Université Paris Cité, CNRS, CNES, Astroparticule et Cosmologie, F-75013 Paris, France\\
$^{2}$Université Paris Cité, CNRS, Astroparticule et Cosmologie, F-75013 Paris, France\\
$^{3}$Université Paris-Saclay, Université Paris Cité, CEA, CNRS, AIM, 91191, Gif-sur-Yvette, France
}

\date{Accepted XXX. Received YYY; in original form ZZZ}

\pubyear{2022}

\begin{document}
\label{firstpage}
\pagerange{\pageref{firstpage}--\pageref{lastpage}}
\maketitle
\begin{abstract}
Accreting supermassive binary black holes (SMBBHs) are potential targets for multi-messenger astronomy as they emit {gravitational waves} (GW) while their environment emits electromagnetic (EM) waves. {In order to get the most out of a joint GW-EM detection we first need to obtain theoretically-predicted EM signals unambiguously linked to BBHs.}
{In that respect,} this is the first of a series of papers \refe{dedicated to} accreting pre-merger BBHs and their associated EM observables.
Here, we \refe{extend} our Numerical Observatory of Violent Accreting systems, {\tt e-NOVAs}, to {any} spacetime.
\refe{Unlike previous studies, almost exclusively focused on the inner regions, we investigated the impact of the BBH on its outer circumbinary disc, located in the radiation (or wave) zone, after implementing an approximate analytical spacetime of spinning, inspiralling BBHs in {\tt e-NOVAs}.}
We follow the formation of a weak spiral structure in disc density \refe{arising from the retardation effects in the radiation zone metric}.
Simulation data are then post-processed with a general-relativistic ray-tracing code incorporating the same BBH spacetime, assuming SMBBH sources.
The density spiral creates a small $({<}1\%)$ but unambiguous modulation of the lightcurve at the semi-orbital period.
This signal, although weak, is fundamentally different from that of an axisymmetric disc around a single BH \refe{\FC{providing} a lower limit on the impact of a BBH on its outer disc}.
This potential difference being found, we study how binary parameters impact this modulation in order to find the optimal case which is a high source inclination of any \FC{binary} mass ratio (from $0.1$ to $1$).
\end{abstract}

\begin{keywords}
black hole physics -- gravitational waves -- accretion, accretion discs
\end{keywords}



\section{Introduction}


A new era for multi-messenger astronomy began with the detection of a post-merger electromagnetic (EM) counterpart to the coalescence of two neutron stars \citep{abbott_gravitational_2017} shortly after the event was detected in gravitational waves (GWs).
Meanwhile, for binary black holes (BBHs), \refe{post-merger detection candidates exist (\citealt{graham_candidate_2020}, \citealt{graham_light_2022}) but no pre-merger counterpart has been detected yet}.
\refe{Actually, a pre-merger} EM-GW co-detection is valuable for many reasons.
\refe{Indeed}, the speed of gravity can be measured and compared to the speed of light \citep{abbott_gravitational_2017} to look for physics beyond General Relativity (GR).
Moreover, the pre-merger EM signal gives us insights on how the plasma behaves in such dynamical spacetimes.
In addition to probing the environment around the new population of astrophysical objects that BBHs unveiled by LIGO/Virgo/Kagra represent, an EM observation allows for new tests of GR in the strong gravity regime.
For these reasons, a pre-merger EM-GW co-detection of BBHs would be a major astrophysical achievement.

The hope for an EM signal relies on the merging system to be gas-rich.
Meanwhile, the BBHs targeted by ground-based GW interferometers are stellar-mass BBHs, whose environment is thought to be cleaned of gas by the supernova explosion leaving the BHs behind.
In that view, the design of the future Laser Interferometer Space Antenna (LISA, \citealt{amaro-seoane_gravitational_2020}), to be launched in the late $2030$'s, could be an asset because GWs emitted by SuperMassive BBHs (SMBBHs), which are more \FC{likely} to be gas-rich systems than stellar-mass BBHs, fall into its frequency range.
Indeed, our current understanding of galaxy evolution is that i) most galaxies host a central SMBH (\citealt{ferrarese_supermassive_2005}, \citealt{gultekin_fundamental_2009}, including the Milky Way, \citealt{gravity_collaboration_detection_2018}) and ii) galaxies grow by repeated mergers (e.g. \citealt{white_core_1978}).
Therefore, the merger of SMBBHs appears as a common phenomenon in the history of the Universe and such an event may occur in a gas-rich environment.
The gas would be provided by the galaxies themselves  during the dynamics of the galaxy merger or directly brought from the SMBBHs accretion environment powering an active galactic nucleus (see e.g. \citealt{alston_super-massive_2022}).
The coalescence of SMBBHs could result in the birth of a single active galactic nucleus, that a pre-merger detection could allow us to witness.
However, the search for a pre-merger EM signal relies on theoretical predictions, which are, to date, uncertain.

A full theory for the accretion onto pre-merger BBHs systems is still in its infancy.
Among the difficulties encountered, one should mention the wide range of spatial scales covered by these systems as they inspiral, in addition to the fluid-related physics to be included
such as magnetic fields and radiation.
At sub-parsec orbital separation, the orbital momentum is mainly evacuated by GWs and GR effects come into play \citep{gold_relativistic_2019}.
There, the spacetime is highly dynamical as it follows the BBH inspiral trajectory together with the continuous emissions of GWs.
Some groups have already incorporated the {temporal dynamics of} {such} a spacetime in their GRMHD code, allowing the study of the circumbinary disc (\citealt{noble_circumbinary_2012}, \citealt{zilhao_resolving_2015}, \citealt{noble_mass-ratio_2021}),
and of the individual accretion structures onto the BHs (\citealt{bowen_relativistic_2017}, \citealt{bowen_quasi-periodic_2018}, \citealt{bowen_quasi-periodicity_2019}) beyond Newtonian gravity (e.g. \citealt{macfadyen_eccentric_2008}).
These involve approximate spacetime constructions (e.g. \citealt{mundim_approximate_2014}, \citealt{ireland_inspiralling_2016}) in which different analytical metrics are stitched together in order to satisfy the approximation they are based on (more details in Sec.~\ref{sec:metric}).
In these constructions, the gravity of the fluid is neglected - which is a reasonable approximation around SMBBHs - and the metric approximations break down as the BHs inspiralling motions become relativistic.
Nonetheless, such an approach is promising because it is less expensive than solving Einstein's equations together with the fluid equations (see e.g. \citealt{noble_circumbinary_2012} and \citealt{giacomazzo_general_2012}).
The system can be evolved on longer timescales and is, therefore, less dependent on its - theoretically-motivated - initial conditions.
Moreover, it can reveal unexpected behaviours that were not captured in Newtonian gravity {such as in} \cite{bowen_relativistic_2017}, who showed that the gravitational potential \refe{is} shallower in GR, \refe{increasing} gas sloshing between the two BHs.
{While all the aspects related to accretion are part of the problem, it does not stop here and the}
photons need to be ray-traced to the observer
to obtain synthetic emission maps and associated diagnostics (spectra, lightcurves).
GR effects (lensing, time delay, gravitational boost...) can be crucial in the vicinity of the BBH and should not be neglected \refe{(e.g. \citealt{casse_rossby_2018} in single black hole discs, for mini-discs around BBHs see e.g. \citealt{pihajoki_general_2018}, \citealt{ingram_self-lensing_2021}, \citealt{davelaar_self-lensing_2022}, \citealt{davelaar_self-lensing_2022-1})}.
{Indeed, some of the aforementioned simulations were ray-traced and the results used to look for potential EM signatures from SMBBHs }\citep{dascoli_electromagnetic_2018}.

{While previous} 
works focused on the vicinity of the BBH{,
in this paper} we take a complementary approach to the identification of EM observables from SMBBHs {of mass ratio higher than 0.1}.
Since inspiralling SMBBHs already emit GWs entering the frequency range of LISA, \refe{we implemented the approximate metric construction presented in \cite{johnson-mcdaniel_conformally_2009}, \cite{ireland_inspiralling_2016} (see Sec.~\ref{sec:metric})} valid for spinning BBHs, \refe{incorporating retardation effects and GW propagation} to investigate whether these could leave an imprint in the outer parts of the circumbinary disc.
\refe{First, the outer part of the disc is colder than the widely-studied inner part and should emit in a different energy band.
Second, such a study will give us an estimate on the minimal impact expected from a BBH on its
\FC{outer} surrounding, compared to a single 
BH. 
Indeed, any other effect produced by the BBH in the inner regions (if an inner disc is present), such as the propagation of spiral density waves up to the outer disc, \FC{may dominate over those presented in this paper if they have time to reach the outer disc before the merger. Indeed, these inner effects propagate at the local radially decreasing sound speed unlike the BBH metric induced effect studied in this paper propagating at the speed of light. In the end all effects} will only add up and introduce additional non-axisymmetries likely to make those systems more easily identifiable. 
}
{To explore this possibility, }we first proceeded to the generalization of our GRMHD code ({\tt GR-AMRVAC}, \citealt{keppens_parallel_2012}, \citealt{casse_impact_2017}) to {any}  
metrics (Sec.~\ref{sec:novas}).
Then, we implemented the approximate metric from \citealt{ireland_inspiralling_2016} in both our GRMHD code and in the GR ray-tracing code {\tt GYOTO} (\citealt{vincent_gyoto_2011}), forming a pipeline dubbed {\tt e-NOVAs} for "extended Numerical Observatory of Violent Accreting systems" \citep{varniere_novas_2018} adapted to the study of accreting compact objects in dynamical spacetimes.
Using this pipeline, we study how \refe{radiation zone \FC{phenomena} affect the density distribution in the outer circumbinary disc} (Sec.~\ref{sec:runs}).
{Then we look at} how
the extent of the impact of \refe{radiation zone effects} on the density translates into actual differences in the observables.
As a second step, we investigate whether these potential differences could be used to  {give some constraint on} 
the mass distribution in the central region (Sec.~\ref{sec:obs}).
\newline

Throughout this paper, we use Greek letters to denote spacetime indices and Latin letters to refer to spatial indices only.
The spacetime metric, denoted $g_{\mu \nu}$, has signature $(-,+,+,+)$.
{We also use the geometric system of units where $\mathrm{G}=\mathrm{c}=1$, leading to characteristic length and time scales} $1 \mathrm{M_\odot}=1.477\mathrm{km}=4.926\times 10^{-6}$~s.

\section{Spacetime decomposition}
\label{sec:nfz}

\refe{While previous numerical efforts including GR effects were mostly focused on the inner region of the circumbinary disc  
(e.g. \citealt{noble_circumbinary_2012} using an approximate metric, and \citealt{farris_binary_2011}, \citealt{farris_binary_2012}, \citealt{gold_accretion_2014-1} solving the BSSN formulation of the Einstein's equations),
we study here 
the impact of \PV{radiation-zone-only} effects, which include retarded potentials associated with the binary black hole source and inherently related to the propagation of gravitational waves (\citealt{blanchet_gravitational_1998}, \citealt{johnson-mcdaniel_conformally_2009}) on the gas-rich environment around BBHs.
}
To perform this study, \refe{those effects} are taken into account via an analytical spacetime. 
We consider a time-varying, analytical, approximate spacetime around an inspiralling BBH system, described by the metric tensor $g_{\mu \nu}$.
At a given time, space is decomposed in several zones, referred to as the Near and Far (or radiation, or wave) Zones (hereafter NZ and FZ, respectively), depending on the distance to the binary system, so that the metric in these zones includes the dominant physical effects as perturbative effects with respect to the flat Minkowski metric.
In each zone the corresponding metrics are derived by making approximations which can depend on the zone itself (e.g. the weak field approximation which depends on the distance to the binary system) or the properties of the binary system (e.g. the orbital velocities of the BHs) and are therefore only approximate solutions to Einstein's equations.
The NZ and FZ share a spatial region in which their respective approximations hold, referred to as the boundary zone.
Using a transition function in the boundary zones, the metrics are stitched together to finally give the global metric $g_{\mu \nu}$.
The transition is performed so as to introduce an error which is smaller than the error already present in the individual metrics.
For each zone, we looked for high-order approximate metrics in the literature.
As we focus on \refe{radiation zone effects on the outer} circumbinary disc, the innermost zone \refe{of our simulation} (here the NZ) is located well outside the event horizon of each BH.

In order to do \refe{so}, one needs to know where the BHs are.
{Indeed,}  the orbital parameters of the binary (the positions, velocities of the BHs...) are necessary to compute their gravitational impact on their surroundings.
The orbital parameters are deduced from several hypotheses regarding the orbital regime of the binary. In particular, we restrict ourselves to quasi-circular orbits with an angular momentum loss mechanism associated to the emission of gravitational waves (see Sec.~\ref{sec:eom}). Hence, this work is adapted to study the final stages before coalescence, {namely a few ${\refe M}/10^6\Msol $~days before merger, for reference }.

\subsection{Determination of the orbital parameters}
\label{sec:eom}

The metric components are functions of the trajectory of the BHs.
We assume it to be a quasi-circular inspiral, with the orbital energy conservation relying on the adiabatic approximation (\citealt{blanchet_gravitational_2014}).
Spins are assumed aligned or anti-aligned with the orbital axis. 
From the adiabatic approximation, the motion is described entirely by the orbital phase $\Phi(t)$.
The change of rate of orbital binding energy $\dot{E}$ is assumed to be strictly equal to the rate of loss of energy by gravitational wave emission $\mathcal{F}$ and change in mass $\dot{M}$ (also called tidal heating) via the equation 
\begin{equation}
    \dot{E}=-\mathcal{F}-\dot{M},
    \label{eq:adiab}
\end{equation}
where the dot superscript indicates the derivative with respect to the {remote observer time.}
The tidal heating $\dot{M}$ is non-zero for spinning BHs only.
Using chain rules, Equation~\ref{eq:adiab} can be rearranged as a function of $\mathrm{d}v/\mathrm{d}t$, where $v\equiv (M \mathrm{d}\Phi/\mathrm{d}t)^{1/3}$ is the {Post-Newtonian (PN)} expansion parameter.
{PN expansions assume that the source is slowly moving, i.e. $(v/\mathrm{c})^2 \ll 1$, and weakly stressed, i.e. $|T^{0i}/T^{00}| \ll 1$ and $|T^{ij}/T^{00}|^{1/2} \ll 1$, where $T^{ij}$ is its momentum-energy tensor.}
One obtains the following set of ordinary differential equations
\begin{equation}
     \diff{v}{t} = -\frac{\mathcal{F}(v)+\dot{M}(v)}{\mathrm{d}E(v)/\mathrm{d}v} ; \quad \diff{\Phi}{t} \equiv \frac{v^3}{M}.
    \label{eq:eomode}
\end{equation}
The energy-balance equation is inverted so that the quantity $\mathrm{d}t/\mathrm{d}v$ is expanded in a series of $v$, then integrated to get $t(v)$.
The expressions of $E(v)$, $\mathcal{F}(v)$ and $\dot{M}(v)$ are given in the ancillary materials of \cite{ajith_ninja-2_2012}.
Finally, the series $t(v)$ is inverted into $v(t)$, then integrated with respect to time to get the phase $\Phi(t)$. \\

We now look for the orbital separation $r_{12}(t)$ as a function of time. 
We expand the right-hand side of
\be
\diff{t}{r_{12}} = -\frac{\mathrm{d}E(r_{12})/\mathrm{d}r_{12}}{ \mathcal{F}(r_{12})+\dot{M}(r_{12}) },
\label{eq:dtdr}
\ee
obtained similarly to Eq.~\ref{eq:eomode},
as a function of $r_{12}$ to integrate it in order to get $t(r_{12})$.
We use Eqs.~229 and 313 of \cite{blanchet_gravitational_2014} for the energy and flux formulas as a function of the parameter \refe{labelled $\gamma$ in \cite{blanchet_gravitational_2014}}, which depends on the orbital separation $r_{12}$.
For the tidal heating, we take Eq.~11 of \cite{alvi_energy_2001}.
Finally, the orbital separation $r_{12}(t)$ is obtained from $t(r_{12})$ and $\mathrm{d}t(r_{12})/\mathrm{d}r_{12}$ via a Newton-Raphson algorithm.
Positions with respect to the center of mass (COM) are obtained from Eq.~5.2 of \cite{blanchet_erratum_2010}. 
Instead of numerically deriving the COM positions with respect to time, we compute the COM velocities using the simplest form for the positions, hence still accounting for the decreasing orbital separation vector with the consistent PN phase.
Expressions used here to derive the phase and orbital separation are valid up to 3.5PN{, i.e. up to} $\mathcal{O} (( v /\mathrm{c})^7)$.

\refe{As will be presented below, 3.5PN is higher-order than the metric implementation. 
A first reason for this is that, as mentioned in \cite{johnson-mcdaniel_conformally_2009}, 
the high-order terms in the orbits can actually become important when computing the BBH metric terms at large distances where the retardation time, when accounted for, becomes large. \FC{Considering a 3.5PN orbit will also help to reach a better depiction of the system dynamics as it will not introduce any significant additional approximation coming from the description of the BHs motion.} 
Furthermore, the part of our code solving the equation of motion is independent from the metric terms: this leaves the possibility for any user, in the future, to implement higher-order metrics when available, and to already dispose of 3.5PN orbits.}

\begin{figure}
	\includegraphics[width=\columnwidth]{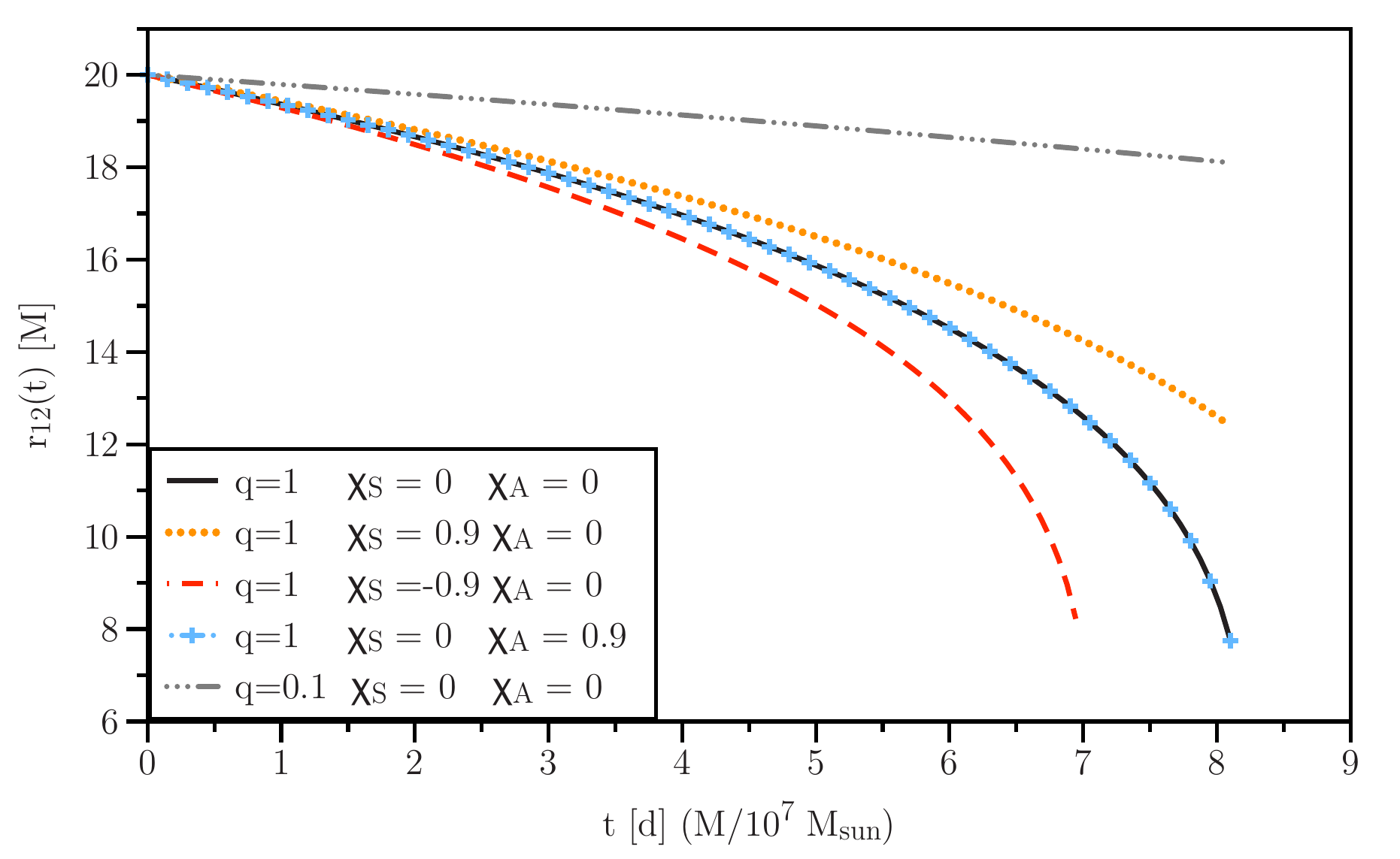}
    \caption{Orbital separation as a function of time expressed in units of days for a $10^7\mathrm{M_\odot}$ BBH, for various values of the binary spin parameters $\chi_S$ and $\chi_A$. Except for the $q=0.1$ system, the time to merger, {from an orbital separation of $20 M$}, is of the order of a few days (for a total mass $M=10^7 \mathrm{M_\odot}$).}
    \label{fig:eom}
\end{figure}

In order to validate our computation of the EOM, we test whether we recover the "orbital hang-up" effect as in \cite{ireland_inspiralling_2016}: a positive spin (with respect to the orbital angular momentum vector) should result in a slower inspiral. Figure~\ref{fig:eom} shows the {evolution of the} orbital separation as a function of time (for an arbitrary initial time {but the same separation}) for different values of the symmetric and antisymmetric binary spin parameters $\chi_\mathrm{S}=(\chi_1+\chi_2)/2$ and $\chi_\mathrm{A}=(\chi_1-\chi_2)/2$, where $\chi_i$ is the spin of the $i-$th BH.
We indeed find that the combination $[\chi_\mathrm{S},\chi_\mathrm{A}]=[0.9,0]$, leads to a slower inspiral than the spinless $\chi_\mathrm{S}=\chi_\mathrm{A}=0$ case.
Oppositely, the combination $[\chi_\mathrm{S},\chi_\mathrm{A}]=[-0.9,0]$ accelerates the inspiral.
Finally, we note that the case $[\chi_\mathrm{S},\chi_\mathrm{A}]=[0,0.9]$ leads to the same trajectory as the spinless case. 

\subsection{Metric components}
\label{sec:metric}
Now that we have the position and velocity of both BHs we can look at the construction of the analytical spacetime \refe{that will be} used in the rest of the paper.
\PV{For that we follow the work by \cite{johnson-mcdaniel_conformally_2009}, and  \cite{ireland_inspiralling_2016} and only present here a succinct explanation of how to obtain their approximate metric.}

Further away than a gravitational wavelength {$\lambda {\sim} 2 \pi / \omega_\mathrm{GW}  {\sim} \pi / \omega_\mathrm{orb} {\sim}\pi (r_{12}^3/M)^{1/2}$} 
, the spacetime metric is the so-called Far Zone metric (FZ, also referred to as the radiation zone), taken from Eq.~6.6 of \cite{johnson-mcdaniel_conformally_2009}.
It derives from the direct integration of the relaxed Einstein equations (also called "DIRE" method) and includes retardation effects, which are crucial in understanding the FZ.
Indeed, the metric components typically vary on the orbital timescale.
As the orbital velocity of the BHs becomes large enough, the retardation effect has a non-negligible impact on the spatial variations of the gravitational field.
A quasi-circular orbit is assumed, i.e. that accelerations and {position vectors} are parallel or anti-parallel, and that velocities are perpendicular to each of these, up to 2.5PN.

At relatively large distances from the binary, i.e. [$r_\mathrm{g}/r=GM/(r \mathrm{c}^2) \ll 1$] and when the motion of the BHs is slow enough, i.e. [$(v/\mathrm{c})^2 \ll 1$], but at a smaller distance than a gravitational wavelength, the Post-Newtonian (PN) approach is valid and useful to give an accurate description of spacetime without  numerically solving Einstein's equations.
In this case, these are solved analytically by developing asymptotic series in powers of these (small) quantities.
The order of the approximation is given by the order of the series, e.g. $(r_\mathrm{g}/r)^2$ gives a "2PN" approximation.
In this paper, we have implemented the NZ metric for spinning BHs (\citealt{blanchet_gravitational_1998}, \citealt{ireland_inspiralling_2016}) including terms up to 2.5PN order (i.e. up to terms $(r_\mathrm{g}/r)^{5/2}$).
Because the PN metric relies on the slow motion approximation, our spacetime is restricted to sufficiently large orbital separations ($r_{12} \approx 8$~M), before motions become relativistic (see Appendix~A of \citealt{noble_mass-ratio_2021} and \citealt{fujita_note_2018}, \citealt{sago_accuracy_2016}, \citealt{yunes_accuracy_2008} in the extreme mass-ratio case).
We also refer the reader to \cite{zilhao_resolving_2015}, who showed that the influence of the PN order on the evolution of a non-magnetized circumbinary disc compares with that of initial conditions further from equilibrium. \\

The global metric is constructed as a weighted sum of the NZ and the FZ metrics
\be
g_{\mu \nu} = (1-f_\mathrm{FZ}) g_{\mu \nu}^\mathrm{NZ} + f_\mathrm{FZ} \, g_{\mu \nu}^\mathrm{FZ}.
\ee
The transition function used between zones is defined as a piecewise function
\citep{mundim_approximate_2014}.
This function returns $0$ {when} $r\leq r_0$, where $r$ is the distance to the center-of-mass of the binary and $r_0$ is a parameter.
For $r_0{<}r{<}r_0+w$, which corresponds to a portion of the boundary zone, it is given by
\begin{equation}
f_\mathrm{FZ}(r,r_0,w,q',s)=
        \frac{1}{2} \left( 1+\mathrm{tanh} \left[ \frac{s}{\pi} \left( \chi (r,r_0,w) - \frac{q'^2}{\chi(r,r_0,w)} \right) \right] \right),
\end{equation}
where $\chi(r,r_0,w)=\mathrm{tanh} \left[ \pi(r-r_0)/(2w) \right]$, and $s,w$ and $q'$ are parameters (let us note that we used $q'$ instead of $q$ as in \citealt{mundim_approximate_2014}, which denotes here the mass ratio).
Finally, this function is strictly equal to $1$ for $r \geq r_0+w$.
Throughout this paper, we follow \cite{mundim_approximate_2014} and take the set of values $s=1.4$, $r_0=\lambda/5$, $w=\lambda$ and $q'=1$, where $\lambda=\pi \sqrt{r^3_{12}/M}$.
The region of validity of each zone is shown in Table~\ref{tab:zones} in the general case and several values of total binary mass.\\

Let us mention several limitations of this construction.
First, as specified above, the slow motion approximation breaks down at orbital separation of about $8$~M.
Second, only the gravitational influence of the binary system is taken into account.
While we intend to use this metric to compute the evolution of a circumbinary disc around a BBH, we will have to restrict ourselves to a non self-gravitating disc, as its contribution to the gravitational field is neglected.
\refe{The self-gravitating region of AGN discs is expected to be further away than ${\sim}2000$ gravitational radii \citep{lodato_self-gravitating_2007}, thus beyond
the region under study here}.
This limitation also sets a lower limit for the mass ratio parameter, which depends on the mass of the disc under study.
Finally, by accounting for orbital shrinkage only due to GW emission (and tidal heating, Eq.~\ref{eq:dtdr}), we neglect the possible torques acting from the disc on the binary.
\refe{Previous studies have shown that the gas-driven torques onto the central BBH are dominated by the negative torques from the individual BH discs (\citealt{tang_orbital_2017}, \citealt{munoz_hydrodynamics_2019}, \citealt{tiede_gas-driven_2020}, \citealt{dittmann_survey_2022}), and could overcome the orbital shrinking due to GWs for $M<10^7\, \mathrm{M_\odot}$ \citep{tang_orbital_2017}.
As we will show in the following, radiation zone effects appear to be dominated by retardation effects associated with the BBH trajectory. 
Hence, any torque acting on the inspiral (or outspiral, e.g. \citealt{dittmann_survey_2022}) would modify as well the (temporal and spatial) periodicity and amplitude of the retardation effects in the radiation zone.
}

\begin{table*}
\caption{Region of validity of each zone of the metric, presented for the general case and for three values of the total binary mass, assuming an orbital separation of $r_{12}=15$~M, in the equal-mass case. $r_A$ stands for the distance to the A-th BH of mass $m_A$, $r$ stands for the distance to the center of mass of the binary.}
\label{tab:zones}
\centering 
\begin{tabular}{c | *2c | *2c | *2c} 
	\hline\hline
	Zone  & \multicolumn{2}{c}{NZ}  & \multicolumn{2}{c}{BZ} & \multicolumn{2}{c}{FZ} \\ \hline 
	{}  & $r_{A}\gg $ & $r_{A} \ll$ & $r \gg$ & $r\ll$ & $r\gg$ & $r\ll$  \\ 
	Theoretical    & $m_A$ & $\lambda$ & $r_{12}$  & $\lambda$ & $r_{12}$ & $\infty$  \\ 
	$M=10^5 \Msol$ [$10^{-5}$~pc]     & $1.6\times10^{-4}$ & $8\times10^{-2}$            & $7.2\times10^{-3}$ & $8\times10^{-2}$ & $7.2\times10^{-3}$ & $\infty$   \\
	$M=10^7 \Msol$ [$10^{-5}$~pc]     & $1.6\times10^{-2}$ & $8$            & $7.2\times10^{-1}$ & $8$ & $7.2\times10^{-1}$ & $\infty$   \\
	$M=10^{10} \Msol$  [$10^{-5}$~pc] & $1.6\times10^{1}$  & $8\times10^{3}$& $7\times10^{2}$  & $8\times10^{3}$ &
	$7\times10^{2}$  & $\infty$  \\  \hline\hline
\end{tabular}
\end{table*}

\subsection{Metric validation}

In order to check that the metric construction presented in Sec.~\ref{sec:nfz} satisfies the Einstein's equations in the vacuum at a satisfactory level of accuracy, 
we compute GR invariants {similarly to the tests done in} \cite{mundim_approximate_2014} and \cite{ireland_inspiralling_2016}.
For the following analysis, we consider equal-mass binaries at orbital separation $r_\mathrm{12}=20$~M located on the $x-$axis, as done in the aforementioned studies, and because it roughly corresponds to the portion of the binary trajectory we are interested in.
The computation is fully numerical, because obtaining the derivatives of the metric components has proven to be computationally prohibitive, despite attempts with the SageManifold tool\footnote{https://sagemanifolds.obspm.fr}.
Spatial and temporal derivatives have been computed using a centered, fourth-order finite difference stencil.
We choose a spacing of $0.8$~M to compute the derivatives in the range $[15,50]$~M and $6.4$~M in the range $[50,1000]$~M for the invariants to reach convergence without being dominated by double precision round-off error.

\begin{figure}
	\includegraphics[width=\columnwidth]{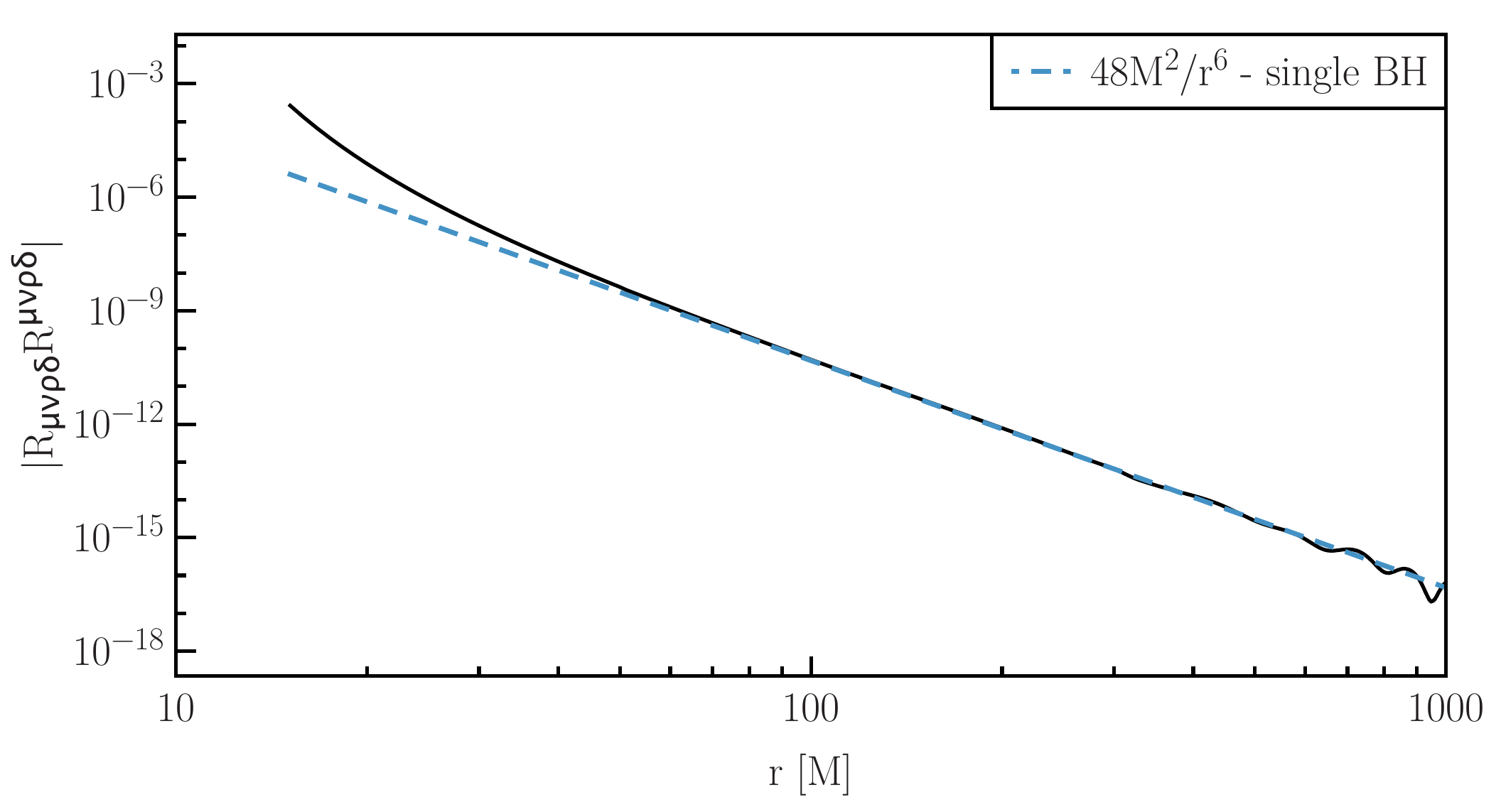}
	\includegraphics[width=\columnwidth]{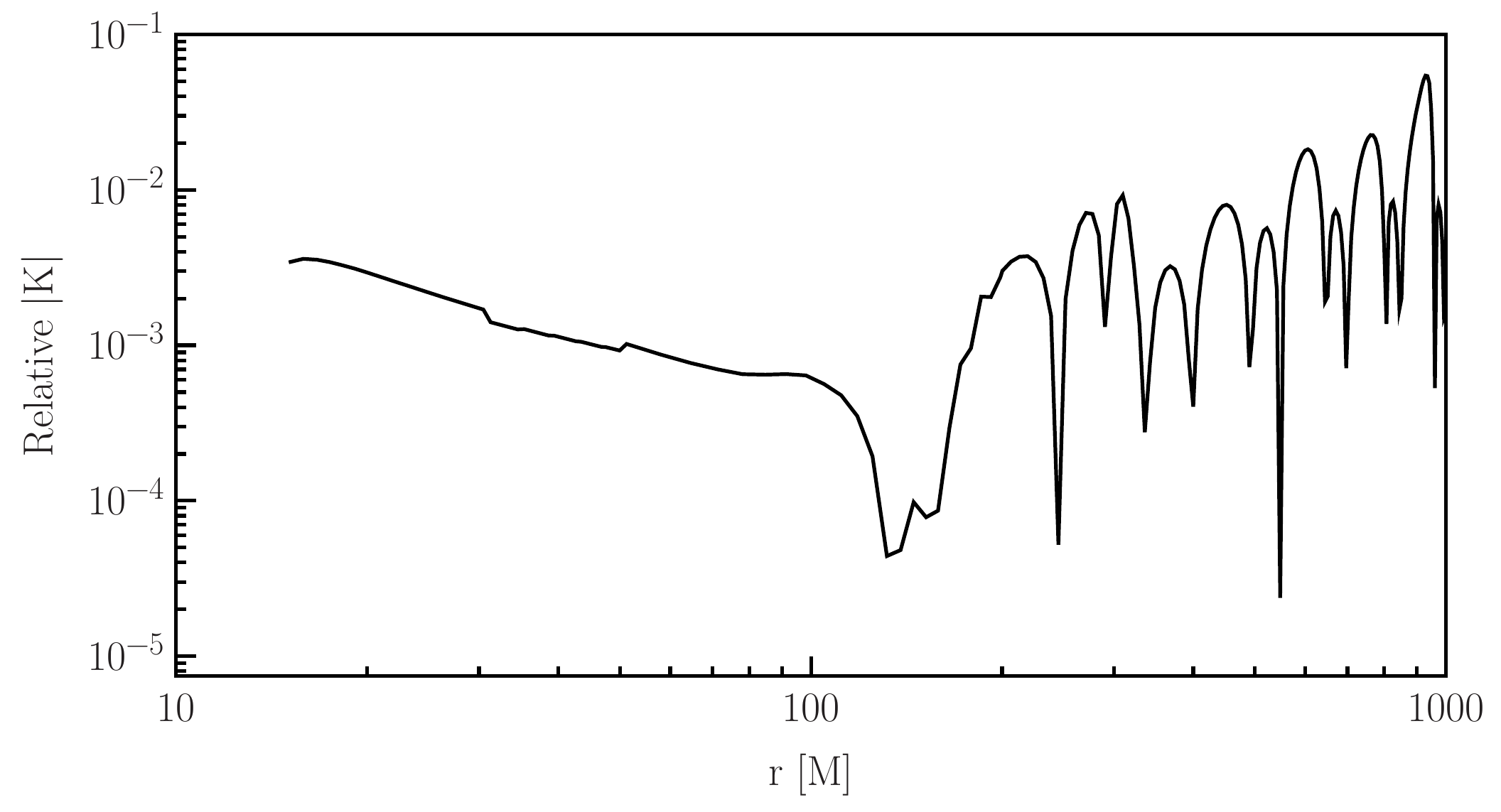}
    \caption{Kretschmann scalar in FNZ (top) and relative Kretschmann (bottom).
    {The BBH has zero spins, equal mass, and an orbital separation of $20$~M.}}
    \label{fig:Kscal}
\end{figure}

First, we compute the Kretschmann {scalar, an} invariant defined as $| R_{\mu \nu \rho \delta} R^{\mu \nu \rho \delta} |$, where $R^{\mu \nu \rho \delta}$ is the Riemann tensor.
Indeed, it can be compared, at least at large distances from the BHs, to the value obtained for a single Schwarzschild BH, $48M^2/r^6$.
The top panel of Fig.~\ref{fig:Kscal} shows the Kretschmann scalar along the $x$-axis
{and how it} decreases as $1/r^6$ at large distances from the BHs, similarly to the slope expected for a single Schwarzschild BH. 

The Kretschmann scalar {deviates from the Schwarzschild value} at small distances from BH1 (located at $r=10$~M), {as expected.
Indeed, in the study of \cite{ireland_inspiralling_2016}, which incorporates the inner zones (i.e. perturbed Kerr metrics), the Kretschmann diverges when approaching the center of BH1 because of the true (physical) singularity there.
The deviation we notice is consistent with this.}
However, we start the computation at $r=15$~M - hence, $5$~M away from BH1 - as the present study focuses on the circumbinary disc.

Second, we compute the relative Kretschmann invariant
, defined as
\be
K_\mathrm{rel} = \left| \frac{2 R_{\mu \nu} R^{\mu \nu } - R^2 }{ R_{\mu \nu \rho \delta} R^{\mu \nu \rho \delta} } \right|,
\ee
where, $R_{\mu \nu}$ and $R$ are the Ricci tensor and scalar, respectively.
Its use as a test for our global metric is justified by the fact that most invariants (i.e. the Ricci scalar) are identically zero.
Hence, they do not provide any comparison scale to measure violations to Eintein's equations \citep{ireland_inspiralling_2016}.
Here, one can directly compare $K_\mathrm{rel}$ to $1$.
The bottom panel of Fig.~\ref{fig:Kscal} displays the value of $K_\mathrm{rel}$ along the $x-$axis.
It can be seen $K_\mathrm{rel}<1$ throughout the domain, confirming that the global metric is a satisfying approximation to a solution of Einstein's equation in the vacuum.

\section{From {\tt NOVAs} to {\tt e-NOVAs} : {general relativistic fluid simulations considering any type of spacetime metric}}
\label{sec:novas}

In \cite{varniere_rossby_2019}, we presented a Numerical Observatory of Violent Accreting systems, or {\tt NOVAs}.
{\tt NOVAs} was composed of the {\tt GR-AMRVAC} code \citep{casse_impact_2017}, a GR version of the finite-volume, {\tt MPI-AMRVAC} code \citep{keppens_parallel_2012}.
{\tt GR-AMRVAC} solves the {general relativistic} equations of magneto-hydrodynamics in Kerr spacetime.
The outputs of simulations {are then} post-processed with the GR code {\tt GYOTO} \citep{vincent_gyoto_2011} to produce synthetic spectra and lightcurves of {gas orbiting around} spinning BHs.
{Here we present  the extended-{\tt NOVAs}  ({\tt e-NOVAs}),  dubbed so as it is 
 a generalization of the {\tt NOVAs} framework beyond the Kerr metric with the aim to include} {any} time-dependent metric.

\subsection{How fundamentally different the BBH metric is from the single BH metric}
\label{sec:bbhvsbh}

{Before going into more details on how we aim to generalize the code to take into account any spacetime metric, let us look at the case we aim to study, namely the 
time-dependent non-axisymmetric spacetime around a BBH system. In particular, what are some of the defining features that depart from the more standard BH metric that is Schwarzschild/Kerr.
{Indeed, those differences motivate the present work.
Because gravitation eventually dominates any other physical effect close to BHs, any difference in the metric affects the fluid dynamics.
Theoretically, it justifies the need to study the accretion onto BBHs as a distinct problem from accretion onto single BHs.
Observationally, differences in the fluid dynamics will play a key role in allowing us to distinguish BBHs from BHs with EM facilities.
These differences will be strenghened as the spacetime does not only influence fluid dynamics but also the propagation of photons towards the observer.
}
}


\begin{figure}
	\includegraphics[width=\columnwidth]{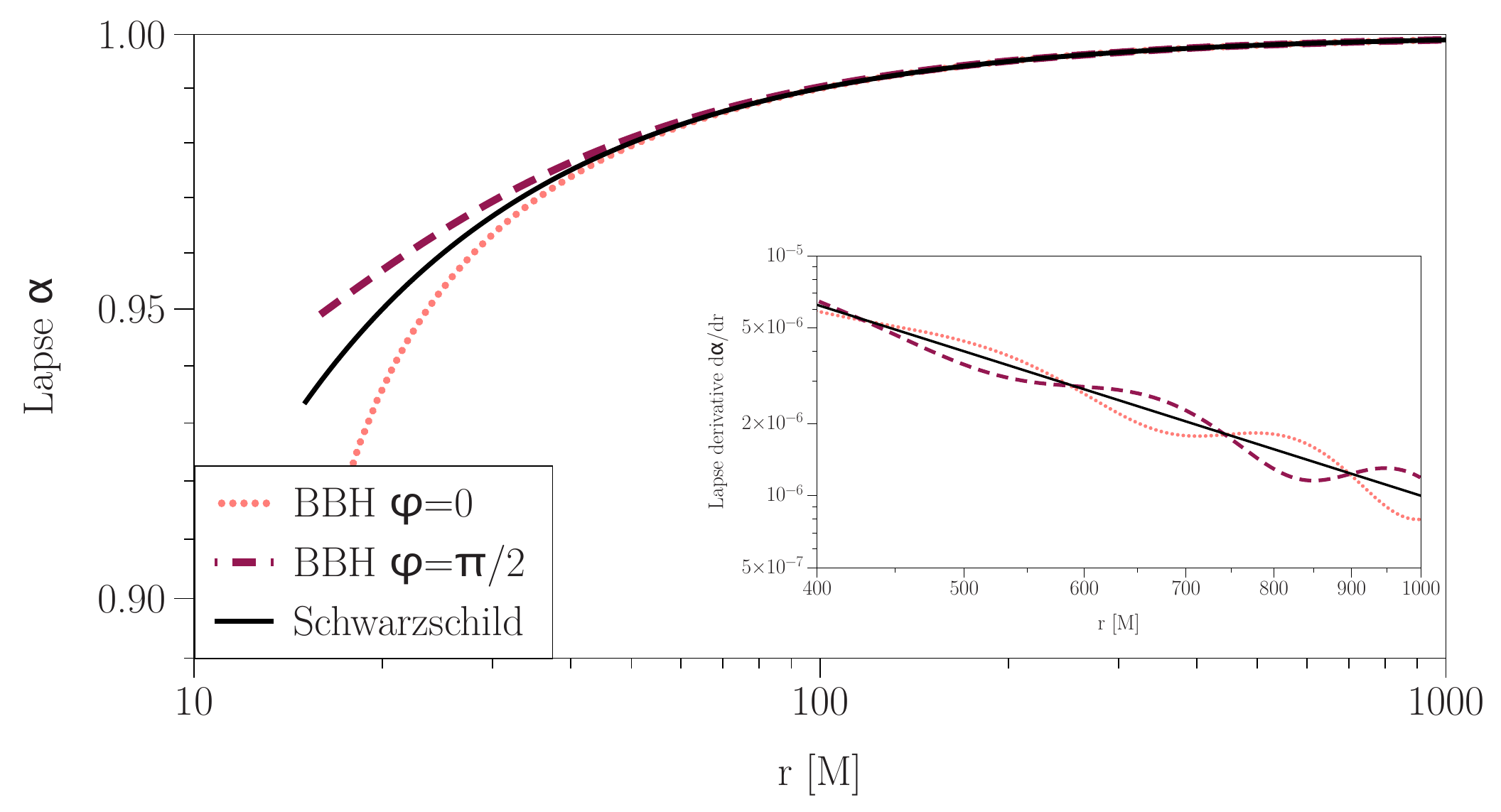}
	\caption{
	Lapse scalar $\alpha$ in the BBH metric for different azimuthal angles against the Schwarzschild metric lapse scalar.
	{The BBH has zero spins, equal mass, and an orbital separation of $20$~M.
	The zoom-in view gives the lapse radial derivative in the FZ region, showing the oscillation.
	}}
    \label{fig:lapse_amrvac}
\end{figure}

In order to illustrate how the BBH metric deviates from axisymmetry and from the Schwarzschild one, we display in 
Fig.~\ref{fig:lapse_amrvac} the radial profile of the lapse scalar \refe{defined as $\alpha=\sqrt{-1/g^{00}}$, which acts as the equivalent of the gravitational potential,} for $\phi=0$ and $\phi=\pi/2$ in both cases.
For comparison, the Schwarzschild metric is computed around a single BH whose mass is equal to the total mass of the binary.
The two BHs are located at $\phi=0$ and $\phi=\pi$, respectively, with orbital separation $20$~M (i.e. in this profile, one BH is located in $r=10$~M).
Close to the binary, the metric non-axisymmetry is visible as the $\phi=0$ and $\phi=\pi/2$ diverge from each other and from the Schwarschild value.
At large distances, the profiles seem to converge towards similar values and therefore a BBH is hardly distinguishable from a single BH, at first sight (relying only on the lapse scalar, see below).
{This trend justifies the initial setup we present in Sec.~\ref{sec:runs}, although another possible approach is to compute a $\phi-$averaged metric to set the initial conditions (see e.g. \citealt{noble_circumbinary_2012}, \citealt{zilhao_resolving_2015}).}

\begin{figure}
	\includegraphics[width=\columnwidth]{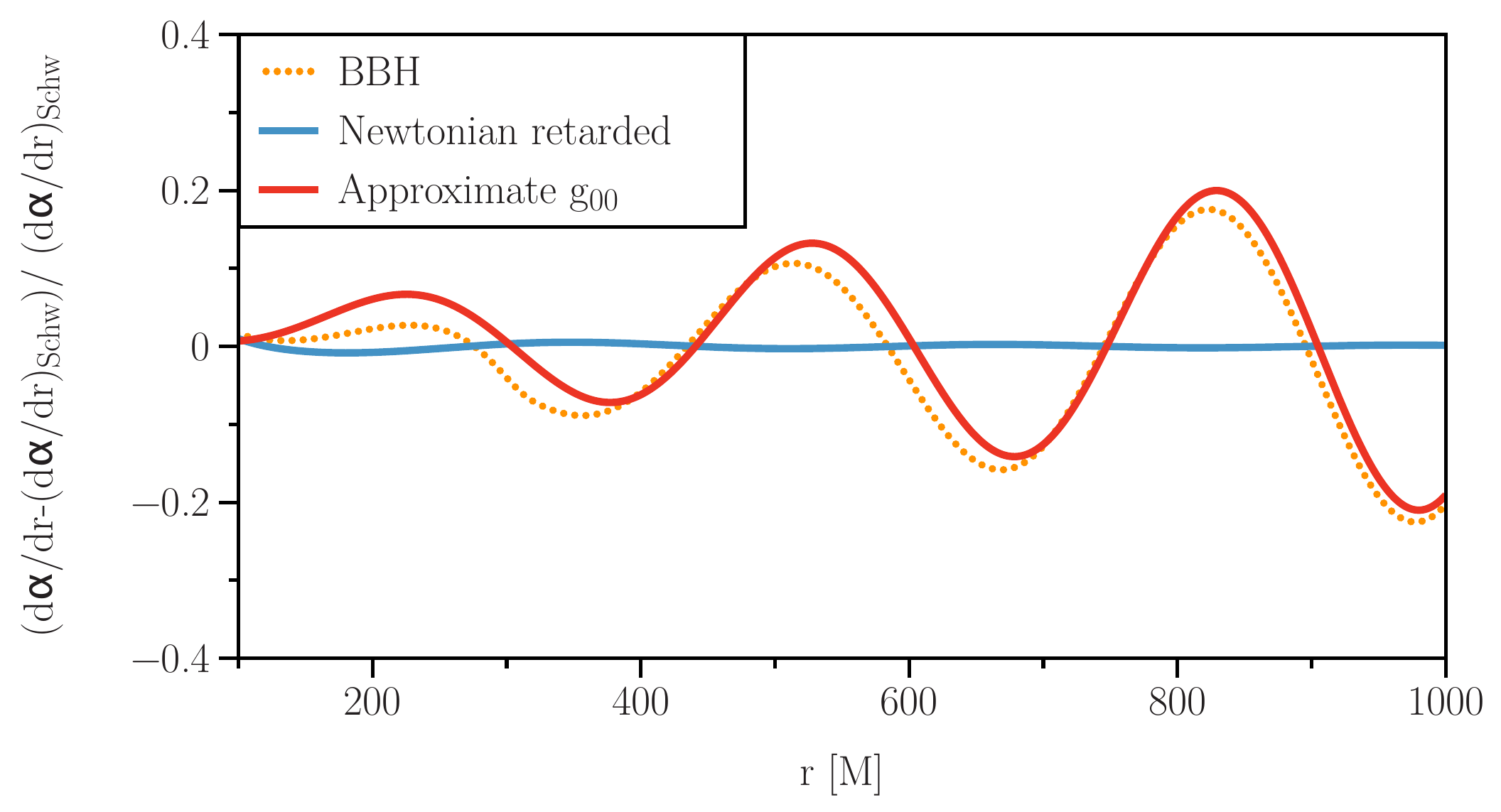}	
	\caption{
	Comparison between the deviation of the lapse derivative in the FZ region with respect to the Schwarzschild metric case, for various BBH modelling (for $\phi=0$). 
	Three  lapse radial derivatives, equivalent to the standard Newtonian gravitational acceleration, are considered: the actual BBH metric (labelled "BBH"), a circular-orbit BBH and its retarded Newtonian potential (labelled "Newtonian retarded", Eq.\ref{Eq:Newt}) and finally an  approximate lapse radial derivative under the circular orbit hypothesis (see main text and Eq.(\ref{Eq:goo})). This figure clearly illustrates that the gravity oscillations affecting the outer region of the BBH circumbinary disc stem from the relativistic nature of the two orbiting BHs.
	}
    \label{fig:lapse_amrvac2}
\end{figure}

{The BBH metric is, nevertheless, fundamentally different from the single BH metric.
In the zoom-in view of 
Fig.~\ref{fig:lapse_amrvac}}, we show the radial derivative of the lapse in the FZ region.
In a Newtonian view, this quantity returns the standard $1/r^2$ gravitational \refe{acceleration.
Here there is a deviation in the BBH case, and this deviation takes the form of a spiral, illustrated by the two oscillations taken at $\phi=0$ and $\phi=\pi/2$ being in opposed phases.}
This Figure illustrates how much the gravitational field described by the BBH metric actually differs from the single BH case even far from the system, and in particular in the zone of interest for our application 
: the FZ, \refe{where  retardation effects and GWs \PV{actively} contribute \PV{to the metric} \citep{johnson-mcdaniel_conformally_2009}.
We show the deviation of the lapse radial derivative with respect to the Schwarzschild case in more details in 
Fig.~\ref{fig:lapse_amrvac2}, where the dotted-lin,e labelled "BBH", is found to consist of a time-dependent oscillation, modulated at the binary period.}
\newline

\refe{In the following, we aim to understand the source of this oscillation.
First, we compute the Newtonian quadrupole potential 
\be
\Phi_\mathrm{Newt} =  \displaystyle\sum_{i=1}^{2} \frac{\mathrm{G \, }m_i}{r_i(t-r)},
\label{Eq:Newt}
\ee
where the distance \FC{to the $i^\mathrm{th}$ BH}, $r_i$, is evaluated at the retarded time $t-r$ (for simplicity, rather than $t-r_i$) and considering a BBH in circular orbit.
We derive the radial gravitational acceleration as $\mathrm{d} \Phi_\mathrm{newt}/\mathrm{dr}$
and represent its deviation with respect to the Schwarzschild value with the blue line.
As clearly visible in the Figure, the Newtonian retarded potential cannot reproduce such an oscillation, and more importantly, it does not introduce any important deviation of the lapse radial derivative compared to the Schwarzschild case. Therefore, the oscillation seen in Fig.~\ref{fig:lapse_amrvac2} comes from \PV{another aspect included in} the BBH metric \PV{not present in the retarded Newtonian potential}.
}

\refe{
\PV{To understand this even further }
we investigate the dominant term of the BBH metric responsible for \PV{those oscillations}.
To do so, we compute the lapse $\alpha=\sqrt{-1/g^{00}}{\approx}\sqrt{-g_{00}}$ by assuming that, because the metric is nearly diagonal, $1/g^{00}{\approx}g_{00}$.
This allows us to compute the lapse for several terms of $g_{00}$ beyond Newtonian order, looking for the term responsible for this oscillation.
We found that (see Eq.~6.6a \citealt{johnson-mcdaniel_conformally_2009})
\be
g_{00}= -1 + \displaystyle\sum_{i=1}^{2} \, 2 \frac{m_i}{r} \left(1+(\rm{v}_i . n)^2\right)
\label{Eq:goo}
\ee
whose associated lapse derivative is displayed as a red line in 
Fig~\ref{fig:lapse_amrvac}, reproduces well the BBH case deviation against Schwarzschild.
In this formula, $\mathbf{v_i}$ is the velocity vector of the $i^{th}$ BH and $\mathbf{n}$ is the normalized position vector of the location we are computing the metric at, \FC{and considering retarded time for all quantities}.
\footnote{For simplicity, we computed $\mathbf{v_i}$ using the circular equation of motion, while the BBH accounts for inspiral; this shows that the inspiral is not responsible for the oscillatory behaviour, at least qualitatively.
Quantitatively, the inspiral will progressively increase $v_i$ as $v_i\varpropto r_{12}^{-1/2}$, the lapse radial derivative and the amplitude of the oscillation along with it, while the oscillation period will smoothly decrease.}
The Schwarschild behaviour of the lapse derivative naturally comes from the term $\displaystyle\sum_{i=1}^{2} 2 \frac{m_i}{r}$ while the oscillation is produced by the term $\displaystyle\sum_{i=1}^{2} 2 \frac{m_i}{r} (\mathbf{v_i} . \mathbf{n})^2$.
This term contributes to the lapse radial derivative only because the retarded time changes the spatial derivatives of $\mathbf{v_i} . \mathbf{n}$ into time derivatives.
Then, the motion of the BBH, periodic in time, translates into spatially-periodic patterns in quantities obtained through spatial derivatives of metric terms.
Unlike the dominant term in the lapse derivative, which goes as $1/r^2$, the contribution of the aforementioned term, responsible for the oscillation, decreases as $1/r$, hence it becomes increasingly important as the radius increases, which is also visible in Fig.~\ref{fig:lapse_amrvac2}.
This term arises from the source retarded potential coming from the diagonal part of the source stress-energy tensor \citep{blanchet_gravitational_1998}. 
As we found the lapse radial derivative of the BBH metric to strongly deviate from that of a single BH metric, with the deviation mainly arising from retardation effects and taking the shape of a spiral, we expect any fluid located in the FZ to be dynamically affected by this.
}
\newline

{The time-dependence of the BBH metric is a major difference between the BH and BBH metrics.
From the numerical point of view, it requires computing the metric at each timestep instead of computing it once and for all.
Moreover, because of the retardation effects in the FZ, one must also compute the metric in each cell in order to obtain the past orbital parameters.
The computational cost is as heavy as the relation between the BHs and time is complex.
Here, a Newton-Raphson inversion is required to obtain the orbital separation as a function of time during the inspiral phase (see Sec.~\ref{sec:eom}).
}

{We discuss here a second aspect of the BBH metric which is not dominant at large distances but may be interesting for future studies incorporating the implementation presented throughout this paper, in particular in the GR ray-tracing step: frame-dragging around BBHs}.
To begin with, the Kerr metric exhibits a non-zero $g_{t\phi}$ component for non-zero spins (and only for non-zero spins).
The radial derivative of the associated shift vector $\beta^{\phi}= \alpha^2 g^{t\phi} $ is responsible for the well-known (axisymmetric) Lense-Thirring effect around spinning BHs.
The BBH metric exhibits such a non-zero and azimuthally-varying component as well (in the center of mass frame), therefore producing a similar but azimuthally-varying effect, {while not needing the BHs to be spinning}.
Interestingly, the azimuthally-averaged amplitude of $\partial_r \beta^\phi$ is comparable in the {maximally rotating} Kerr case and in the BBH case for an orbital separation of typically $10$~M, at comparable distances, showing that the amplitude of the effect is not minor.
This is an example of the interesting physical features arising from the BBH metric compared to the single BH metric.
{Finally, unlike for the Kerr BH, the frame-dragging effect exhibited by a BBH is not axisymmetric}.

{Hence, we have shown here how specific features from the BBH metric will alter either the gas dynamics or the trajectory of photons, or both, with respect to the Kerr BH case.}
For a direct one-to-one component comparison between the NZ metric and the Kerr metric, we refer the reader to \citealt{zilhao_resolving_2015}.

\subsection{Implementation of {\tt e-NOVAs}}
\subsubsection{{Hydrodynamics} in a non-stationary spacetime}
 
{The first step  in studying BBH systems is to}  generalize {\tt GR-AMRVAC} to {take into account} {fully general} time-dependent metrics. {In order to do that we} slightly modified {the definition of our set} of conservative variables as well as the frame in which the corresponding Riemann problem is solved.\\

The set of equations to be solved includes the local conservation of baryon density and the local conservation of {momentum} in a time-dependent background metric, namely{
\be
\beal
\partial_t ({\cal D}) + \partial_j \left[  {\cal D} \left( \alpha {\rm v}^j - \beta^j \right) \right] =\ &0& ,\\
\partial_t ({\cal S}_i) + \partial_j \left(  \left[ {\cal S}_i (\alpha {\rm v}^j - \beta^j) + \alpha{\cal P}\delta_i^j \right] \right) &=& \\
-(W{\cal D}+{\cal P}(W^2-1))\partial_i \alpha + \frac{\alpha}{2} \left( {\cal S}^j {\rm v}^k + {\cal P}\gamma^{jk}\right)\partial_i \gamma_{jk} + {\cal S}_j \partial_i \beta^j &&,
\label{eq:eqs}
\eeal
\ee
where $\alpha$ stands for the lapse function, $\gamma_{ij}$ for the spatial metric tensor components and $\beta^i$ for the shift vector components. The conservative variables in this fully general framework are the relativistic density ${\cal D}=\sqrt{\gamma}W^2\rho$ and the relativistic momentum ${\cal S}_i=W({\cal D}+W{\cal P}){\rm v}_i$. In the previous definitions, $\sqrt{\gamma}$ stands for the square root of the spatial metric determinant, $\rho$ is the rest density of the gas and ${\rm v}^i$ its contravariant velocity vector. The Lorentz factor is noted as $W=(1-g_{ij}{\rm v}^i{\rm v}^j)^{-1/2}$ while the local gas pressure $P$ is linked to the pressure ${\cal P}$ measured in the observer frame ${\cal P}=\sqrt{\gamma}P$. Finally $\delta_i^j$ stands for the Kronecker symbol. }
As can be seen from the Eq.~\ref{eq:eqs} above, generalizing the code to {non-stationary metrics, the spatial metric 
determinant now enters the definition of conservative variables, thus taking into account  effects of the non-stationarity of the gravity upon the gas properties.}

{Stepping into fully general metrics also affects the way the (magneto-)hydrodynamical solver addresses the Riemann problem.
Indeed, such a problem can be solved easily in any direction in a flat spacetime, namely in a locally Minkowski frame. }
To do so, we build a new basis $\{ \mathbf{e}_{(\hat{\mu})} \}$ ensuring that the inner product gives $ \mathbf{e}_{(\hat{\mu})}  \cdot  \mathbf{e}_{(\hat{\nu})}  = \eta_{\hat{\mu}\hat{\nu}}$, where $\eta_{\hat{\mu}\hat{\nu}}$ is the Minkowski metric  (more details in section 3.3 of \citealt{white_extension_2016}).
The corresponding frame transformations are performed using the matrices $M^\mu_{\hat{\nu}}$ and $M^{\hat{\mu}}_\nu$  provided in Appendix~\ref{app:fido}.

\subsubsection{Ray-tracing {in a non-stationary spacetime}}

{Once we have the gas dynamics around the BBH, we still need to ray-trace the emission back to the observer if we want to obtain observables. 
In order to do that we have implemented  the same metric as in {\tt GR-AMRVAC} in  {\tt GYOTO}.
While this was the first time a non-stationary spacetime was added in it, the necessary infrastructure was already present in the general code and adding out analytical metrics was} 
relatively straightforward {after the work done to implement it in {\tt GR-AMRVAC}}.

\begin{figure}
	\includegraphics[width=\columnwidth]{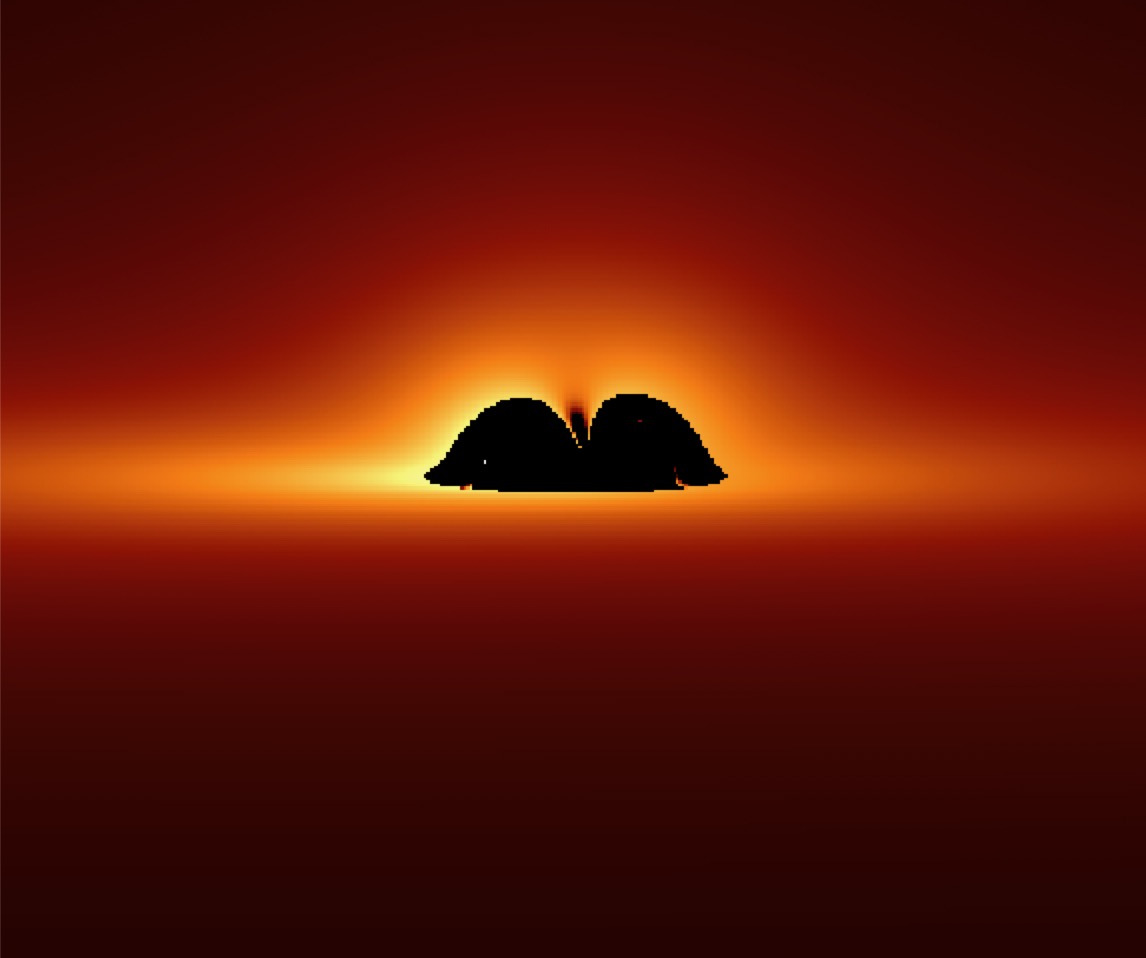}
	\caption{Emission map of a simple axisymmetric disc model incorporating the FZ metric in {\tt GYOTO}. The BBH has $q=1$ and individual spins $0$ and $1$.
	The source inclination angle is $85\deg$ (close to edge-on).
	Around a single static BH with an axisymmetric disc, the image would be distorted by various time-independent effects (lensing, gravitational boost and the Lense-Thirring effect).
	Here, the distorsion of the disc image is time-dependent because it follows the BBH trajectory.}
    \label{fig:em_map_bbh}
\end{figure}

{After the implementation we} checked that, far from the BBH {we obtain similar images compared} to a unique Kerr BH of mass equal to the total binary mass {as is expected from the metric (see Sec.\ref{sec:metric})}. 
Meanwhile, close to the BBH, rays {show a strong deviation from the single BH case}
and {illustrate} the influence that {taking into account} the BBH can have on the ray-tracing step{, and therefore on the lightcurve and energy spectrum computed with it}.
The illustration of such distorsions is given in Fig.~\ref{fig:em_map_bbh}.

Moreover, the spatial variations of the lapse scalar $\alpha$ shown in Fig.~\ref{fig:lapse_amrvac}, as well as its non-axisymmetry, suggest interesting effects associated with the retardation of light rays coming close to the BBH, {as well as non-axisymmetric frame-dragging at small distances from the BHs, as mentioned previously.
Those effects have the potential to affect not only the emission at a given local time but also the timing features of the source, since any non-axisymmetry associated to the BBH evolves on the binary period.}
\newline 

\refe{Once we have the metric in which the photon evolves, {\tt GYOTO} ray-traces the photon backward in time from the observer until the accretion disc is reached and uses the temperature map produced by the GR simulation to compute the blackbody emission from it.}

\refe{\noindent The temperature map for each cases is produced using $T(t,r,\phi) = T_\mathrm{in} (\rho(t,r,\phi)/\rho_\mathrm{in})^{\gamma'}$ where $\rho_\mathrm{in}$ is the density at the inner edge of the simulation box a $t=0$, $T_\mathrm{in}$ is the associated temperature and $\gamma'$ is the adiabatic index (see Sec.~\ref{sec:setup}).}
Since we focus our study on SMBHs, we use an observational template taken from \cite{lin_38_2013}.
They fitted the energy spectrum of a $M=10^5\Msol$ BH with a blackbody model, deriving a temperature at the innermost stable circular orbit of about $0.1$~keV. 
These values, along with a distance of ${\sim}500$~Mpc, serve as a reference for the cgs fluxes to be computed.
As we want to span the mass range from $10^5$ to $10^{10}\Msol$, covering a wide range of masses of SMBBHs, 
\refe{we decided to use the mass scaling of the temperature presented in the standard disc model of \cite{shakura_black_1973}.
This gives $T_\mathrm{in} \propto (M/\mathrm{M}_\odot) ^{(-1/4)}$ (see also, e.g. \citealt{rees_black_1984}).}

\section{Impact of \refe{radiation zone effects} on {the} circumbinary disc}
\label{sec:runs}

{As a first example of what {\tt e-NOVAs} could do, we decided to focus on the circumbinary disc, especially in the Far Zone that is much less studied than region closer to both BHs.} 
{While the existence of a such circumbinary disc has not been confirmed observationally, it}  could be formed when part of each BH's initial disc material has been expelled via gravitational interactions 
(e.g. tidal disc truncation, \citealt{papaloizou_tidal_1977}, \citealt{paczynski_model_1977}) to circularize around the system as a circumbinary disc.
In the case of SMBBHs, on which we focus here, the initial material (if any) is expected to come from their host galaxy-related gas, the same that powers luminous activity in galactic centers, and follow them as they inspiral \citep{haiman_identifying_2009}. {It could also come from the partial or total tidal disruption event of a star originally orbiting close to one of the black-holes, the orbit of which was disturbed by the approaching secondary BH (\citealt{coughlin_influence_2019}, \citealt{shu_x-ray_2020}, \citealt{huang_possible_2021}).}
\newline

{The FZ is characterized by the \refe{retardation effects associated with the BBH potentials \citep{blanchet_gravitational_1998} consistent with the} propagation of the GW produced by the inspiralling BBH. Thanks to {\tt e-NOVAs} we can study \refe{how the radiation zone metric affects} the circumbinary disc and simultaneously explore how \refe{this impacts} standard observables such as the energy spectrum and lightcurve of the system.}


\subsection{Circumbinary disc setup}
\label{sec:setup}

\refe{Simulations are two-dimensional, in the orbital plane of the binary.}
As the FZ metric asymptotically converges towards a Schwarzschild metric at large distances, we initialize {our circumbinary disc as} a disc at equilibrium in a Schwarzschild metric ({with a}
 BH of mass equal to the total mass of the binary and located in its center of mass, as done in e.g. \citealt{farris_binary_2011}) 
 {which will be close to equilibrium, at least in the far region in which we are interested in.}
The equilibrium is given by the (GR version of the) thermal pressure gradients and the centrifugal acceleration in an equivalent Schwarzschild metric.
The initial gas surface density profile given by:

\be
\rho(r) = 0.5 \rho_0 \left( 1 - \tanh{\left( \frac{r-2500}{500} \right) } \right) r^{-3/4} + \rho_\mathrm{min},
\ee
where $\rho_0=10$.
The $\tanh$ function is used to set a disc outer edge, and $\rho_\mathrm{min}=2 \times 10^{-4}$ prevents the density from becoming too small there. Pressure is computed as $p=p_0 \rho^{\gamma'}$, where $\gamma'=5/3$ is the adiabatic index with $p_0=1.8\times 10^{-4}$.
The resulting disc aspect ratio, as computed in Newtonian gravity, is $H/r=0.1-0.25$.
The initial azimuthal velocity is set so as to impose radial equilibrium in an equivalent  Schwarzschild   metric (see e.g. \citealt{casse_impact_2017})
\be
v^\phi = \mathrm{c} \frac{ - \partial_r \beta^\phi \pm \sqrt{ (\partial_r \beta^\phi)^2 + 2 \alpha \gamma^{\phi \phi} \partial_r \gamma_{\phi \phi} (\partial_r \alpha + \alpha \frac{\partial_r P}{\rho h W^2}) } }{\alpha \gamma^{\phi \phi} \partial_r \gamma_{\phi \phi} },
\ee
as in \citealt{farris_binary_2011}. 
\refe{Neglecting the disc self-gravity restricts the value of the surface density normalization, $\rho_\mathrm{0}$.
Self-gravity is negligible as long as $\rho/H \ll M/r^3$ \citep{frank_accretion_2002}. This sets $\rho_\mathrm{0} \ll 10^{13} (M/10^5\mathrm{M_\odot})^{-2} \mathrm{g \, cm^{-2}}$, i.e. $\rho \ll 3\times 10^{10} \mathrm{g \, cm^{-2}} (M/10^5\mathrm{M_\odot})^{-2} \mathrm{g \, cm^{-2}}$ at $r=2100$~M.}
\newline

The computational grid extends radially from $18.75$~M to $5000$~M, and from $0$ to $2\pi$ azimuthally. 
The resolution is $n_r \times n_\phi = 476 \times 140$ with logarithmic spacing in the radial direction resulting in the finest resolution {of $4$~M in the inner region of the FZ and $20$~M at its outer edge, which 
always gives a resolution better than about $20$ cells per gravitational wavelength.}
We checked that running a higher resolution simulation\refe{s (see Sec.~\ref{sec:impact})} leads to qualitatively \refe{and quantitatively} similar results.
The inner and outer boundary conditions are null-gradient on primitive variables, i.e. the gas density $\rho$ and the contravariant velocity vector components $v^i$. {Outflow conditions are imposed at the outer boundary of the simulation}.
\newline

{We are using} the global metric (NZ, FZ, and transition region, see Sec.~\ref{sec:metric}) {with the}
NZ  used to provide self-consistent boundaries to the region of interest, i.e. the region covered by the FZ. This means that while the \refe{computational grid is more extended, }we focus
on the region covering $[1000,2100]$~M in radius where there is no contamination from effects related to the NZ as they have not had time to reach the region of interest. 
\refe{By doing so, we are assuming that i) either the effects we will report are not dominated by \FC{any fluid perturbations originating from the NZ and slowly propagating} out to the FZ \refee{- indeed, the time it takes for spiral waves to reach $r=1000$~M at the local sound speed is ${\approx}6\times 10^4$~M, which is between five and twelve times the inspiral time in the following simulations, depending on the mass ratio -}, or ii) that there is no disc in the NZ, as could be the case during the formation of the circumbinary disc or considering the more general case of diluted gas being located in the radiation zone.
In any case, this problem has not been addressed in the literature and we propose to study it as an application case of our pipeline.
If the effects we report here are eventually dominated by other BBH-related ones \FC{(but probably not acting at the same frequency)}, then those will add up and \FC{our results will thus stand as} a lower limit on the influence of a BBH on its outer circumbinary disc.
Future studies, in this series of papers, will focus on the inner regions of the circumbinary disc \citep{mignon-risse_origin_2022}.}

\subsection{Impact of \refe{radiation zone effects} on the density distribution}
\label{sec:impact}

In order to study the impact of \refe{radiation zone effects on the disc} we decided to study three distinct mass ratios. First we will look at the $q=1$ case \refe{as a reference case.}
As a second case we will study the $q=0.3$ case as it is typical of the sources LISA 
could detect \citep{colpi_massive_2014} and therefore has a good potential to be observed. Lastly, we will look at an even smaller mass ratio of $q=0.1$ to see how different is its impact.
Indeed, unequal mass ratios are naturally expected in the merger of SMBBHs, unless some mechanism(s) drives preferential accretion onto the least massive component of the binary (e.g., \citealt{farris_binary_2014}, \citealt{duffell_circumbinary_2020}; see also \citealt{bate_predicting_2000} and \citealt{mignon-risse_collapse_2021-1}, \citealt{mignon-risse_magnetic_2023} in the protostellar context), but this question remains open \citep{siwek_preferential_2022}.

In all the simulations we follow the inspiral motion of the binary from an orbital separation $r_{12}(t=0)=15$~M down to $8$~M, where the PN approximation breaks down. 
\refe{Let us note, however, that this portion of the BH trajectory is arbitrary and that these simulations are not limited by their computational cost.
Starting simulations of circumbinary discs with a larger orbital separation is feasible but it would become increasingly relevant to take into account the potential individual accretion structures feeding the black holes, which is not our concern here. In addition, the FZ would be located further away (see Table~\ref{tab:zones}), so the simulation box should be enlarged and the absence of self-gravity in our model may become a concern.}
The computational time is equivalent to $t {\sim} 4000 \mathrm{M/c} \ {\sim}10P_\mathrm{orb}$ for $q=1$, where $P_\mathrm{orb}$ is the initial orbital period; as the BBH orbit shrinks, the orbital period decreases.

\begin{figure}
	\includegraphics[width=\columnwidth]{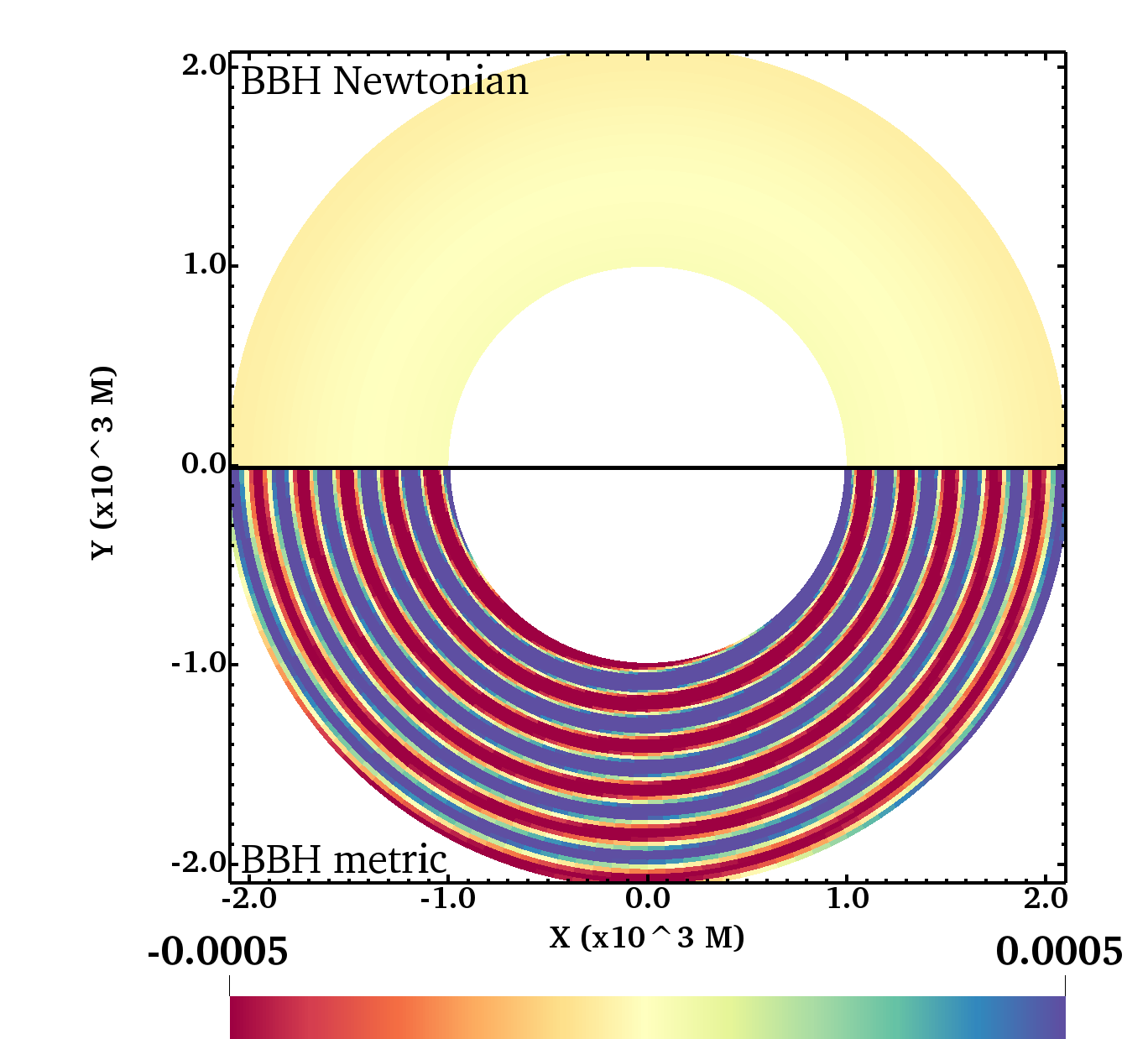}
    \caption{Map of the density deviation, $(\rho(t)-\rho(t=0))/\rho_\mathrm{Schw}$, where $\rho_\mathrm{Schw}$ is the single BH case density, in the case of equal-mass BHs in the Newtonian case (top) and in the BBH metric case (bottom), at $t=4000 \mathrm{M}$.
    In the Newtonian case an axisymmetric relaxation wave propagates outwards.
    In the BBH metric case, any non-axisymmetry arises as a reaction of the fluid being in the BBH metric rather than in a Kerr metric.
    Distances are in units of $M$.}
    \label{fig:FNZrho}
\end{figure}

\begin{figure}
	\includegraphics[width=\columnwidth]{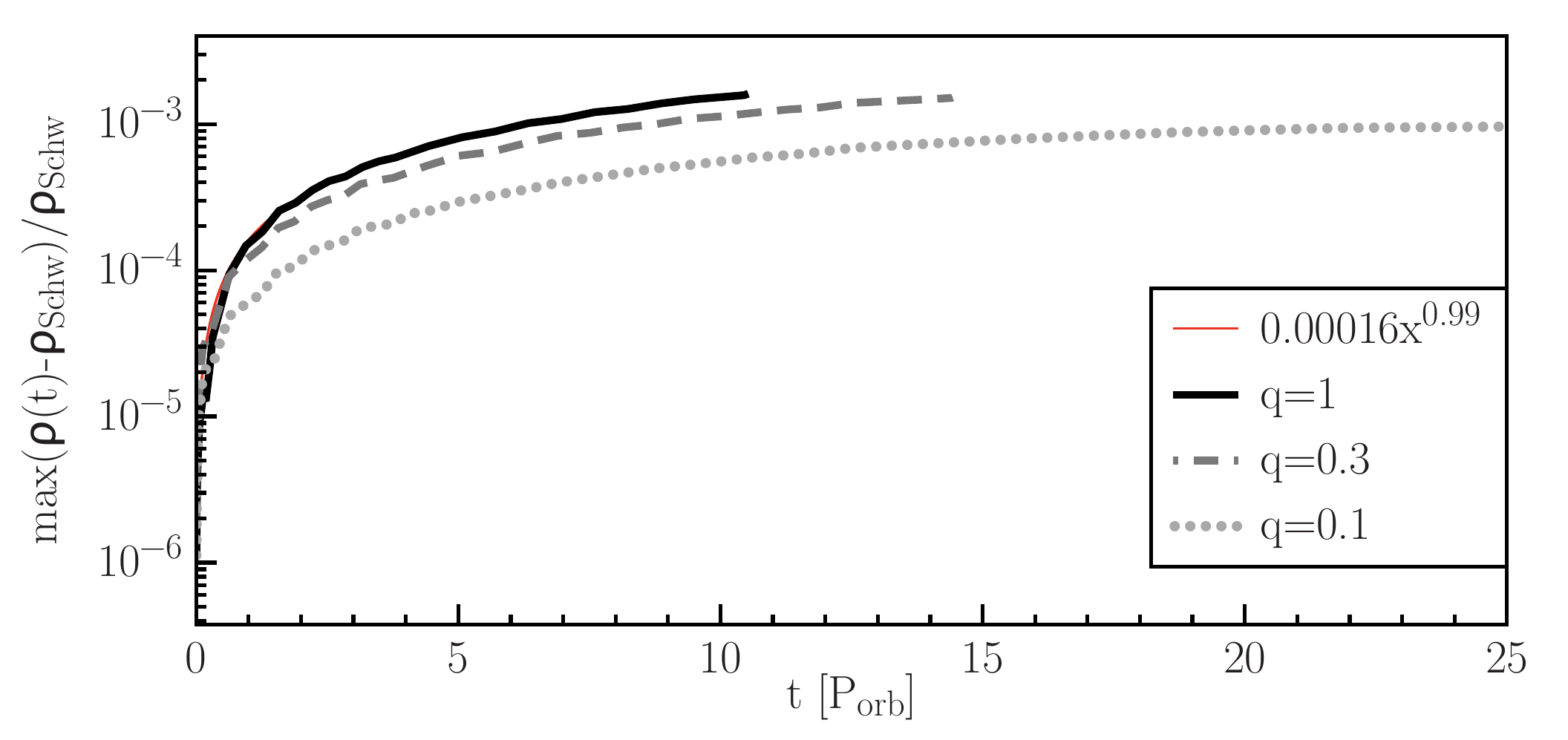}
    \caption{{Evolution of the} amplitude of the spiral structure, shown as $\max(\rho(t)-\rho_\mathrm{Schw})/\rho_\mathrm{Schw}$, where $\rho_\mathrm{Schw}$ is the single BH case density, as a function of time until an orbital separation of $8$~M is reached. Several values of the mass ratio parameter $q$ are explored: $q=1$ (black curve), $q=0.3$ (dark gray dashed curve), $q=0.1$ (light gray dotted curve). The red curve is a power law fit to the $q=1$ case between $2$ and $10 P_\mathrm{orb}$.
    The plot covers $10$, $14$, and $24$ orbital periods for $q=1,0.3,0.1$, respectively.
    The time to merger is a few days for $M=10^7\Msol$ (more than ten days for $q=0.1$).}
    \label{fig:FNZrhot}
\end{figure}

Figure~\ref{fig:FNZrho} shows the map of the density deviation with respect to the single BH case, with the BBH modelled using Newtonian gravity (upper panel) and using the BBH approximate metric (bottom panel), at $t=4000$~M. 
In the Newtonian case the density is axisymmetric 
\textit{at these distances} from the BBH.
It weakly deviates from the single BH case because of a relaxation wave propagating outwards at local sound speed, much smaller than the speed of light, with maximal density deviation $5\times 10^{-5}$ in absolute value.
Meanwhile, a spiral in density deviation is clearly visible in the BBH case and has a much larger amplitude than the relaxation wave observed in the Newtonian case.
This spiral is a consequence of the radiation zone effects, including retardation effects acting on the outer disc (as presented in Sec.~\ref{sec:bbhvsbh}), as they cannot be reproduced \FC{neither} in Newtonian gravity nor with the BBH metric without those retardation effects.

While seeing an impact of radiation zone effects on the disc is interesting, we need to quantify it, especially as function of time.
Figure~\ref{fig:FNZrhot} shows the time evolution of
the maximal value of $(\rho(t)-\rho_\mathrm{Schw})/\rho_\mathrm{Schw}$, where $\rho_\mathrm{Schw}$ is the equivalent single BH case density, for various mass ratios of interest.  
We see that the amplitude of the spiral structure increases with time, regardless of the mass ratio.
Nevertheless, the effect is rather weak, as expected: this quantity is dimensionless and is $\lesssim 10^{-3}$.
In order to quantify its temporal evolution, we fit it with a power law function for $q=1$, shown by the red curve.
The power law index is close to unity ($0.99$), so the amplitude increase is roughly proportional with time. It is noteworthy that this is also the case for $q=0.3$ and $q=0.1$.
A natural explanation would be the nearly constant forcing exerted by the metric.
This may sound surprising as GR effects should gain amplitude as the BBH approaches merger: for instance,
the amplitude of the oscillation reported in the lapse radial derivative increases during the inspiral.
However, this is moderated by two effects.
First, the inspiral EOM dictates a separation shrinking as $r_{12}\varpropto (t-t_\mathrm{merger})^{1/4}$, so the binary is at nearly constant orbital separation during most of the computational time.
Second, the gas in the radiation zone is located beyond $1000$~M from the binary, hence at the time the simulation stops because the orbital separation reaches $8$~M, the gas actually reacts to the influence of the BBH at a larger separation (${\approx}11$~M for $q=1$, see Fig.~\ref{fig:eom}). 
Therefore, the metric forcing, as felt by the gas located in the far zone, is nearly constant over the BBH trajectory portion we have studied.

Finally, we also find the amplitude to increase with $q$, and therefore to be the largest in the equal-mass case.
As a side note, the difference in run duration in Fig.~\ref{fig:FNZrhot} comes from the impact of the mass ratio on the inspiral time.
Indeed, our simulations cover the inspiral phase from $r_{12}=15$~M to $r_{12}=8$~M, which takes more time for smaller $q$ values (see Fig.~\ref{fig:eom}).

\refe{Let us also note that we performed higher resolution runs with $n_r \times n_\phi = 952 \times 280$ (labelled {\tt HR} run), and $n_r \times n_\phi = 1428 \times 420$ ({\tt VHR} run), i.e. a twice and three times as better resolution in both directions, respectively.
We computed the relative difference in the aforementioned quantity when increasing the resolution.
From the fiducial run to run {\tt HR}, the relative difference asymptotically decreases with time and remains below $0.5\%$ after ${\sim}2$ orbits.
From run {\tt HR} to {\tt VHR}, it remains below $0.5\%$ after less than $1$ orbit.
Hence, convergence is achieved.}
\newline

We found that \refe{radiation zone effects associated with} inspiralling BBHs\refe{, and dominated by retarded potential contributions,} produce non-axisymmetries in their circumbinary disc in the form of a spiral structure.
{While we found this structure to have a small, but growing, amplitude it is worth looking at its electromagnetic emission to see if the departure from axisymmetry would have an impact on it as well.}

\section{Is there any observable imprint from radiation zone effects on the circumbinary disc?}
\label{sec:obs}

In this section, we look at the impact of the spiral structure related to the \refe{radiation zone effects on} a SMBBH circumbinary disc on the EM observables.
To create synthetic lightcurves and spectra, we post-processed the simulations with the {\tt GYOTO} code {under the assumption that the} circumbinary disc emits thermal blackbody radiation.
{While the direct impact of the metric on the disc was small, as the photon will be traveling through the same spacetime on their way to the observer, they will also be affected by the 
metric which could boost, slightly, the effect.}

\subsection{Energy spectrum for various BBH masses}

\begin{figure}
	\includegraphics[width=\columnwidth]{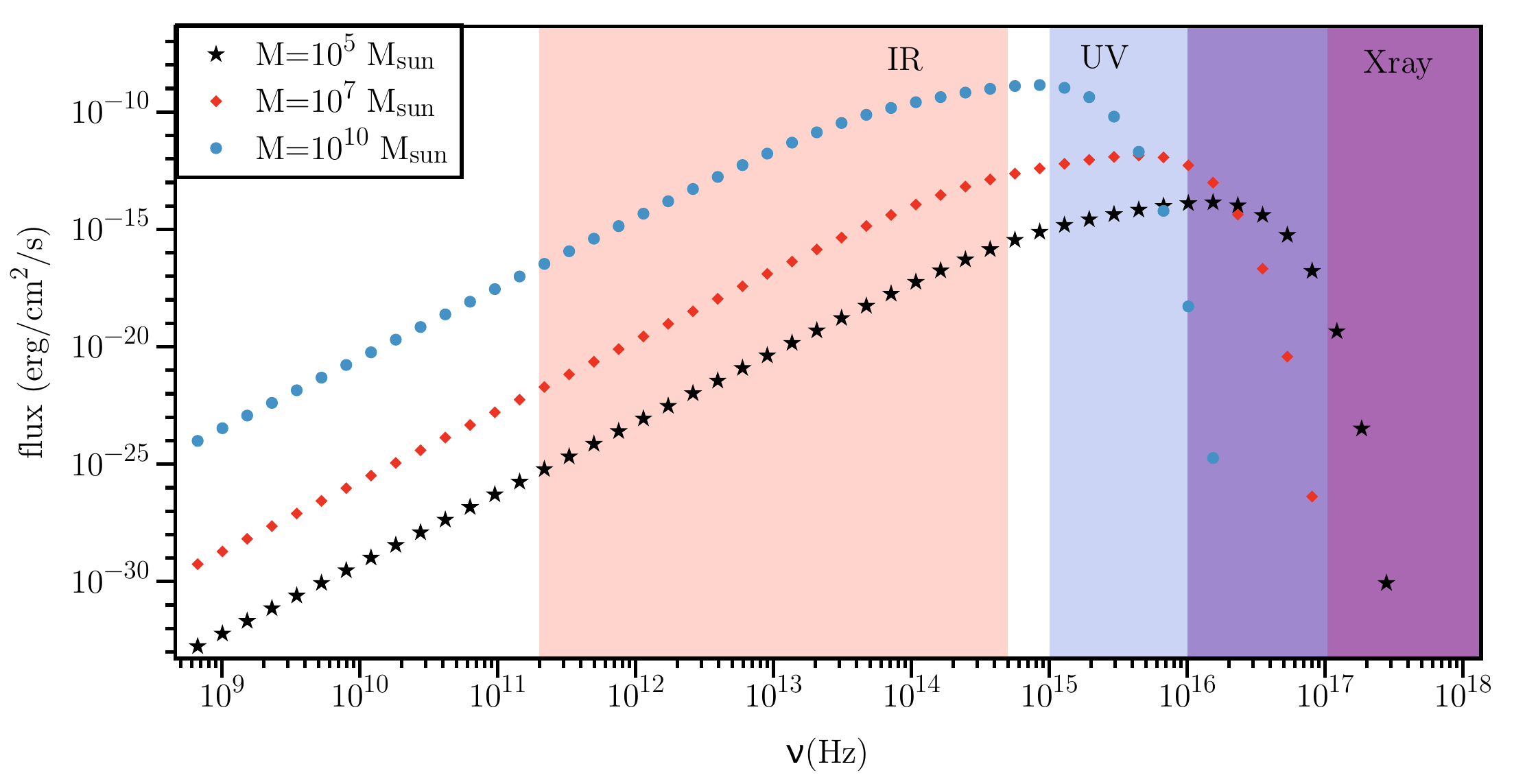}
    \caption{Energy spectrum at $t_\mathrm{8M}$, for $q=1$ ranging BBH masses from $10^5$ to $10^{10}\Msol$\refe{, at a distance of $500$~Mpc}.
    To illustrate the capabilities of {\tt e-NOVAs}, the displayed spectra were computed for spinning BBHs with spin parameter $\chi_S=0.9$; the quantitative difference with respect to $\chi_S=0$ is, however, very small \citep{mignon-risse_gravitational_2022}.}
    \label{fig:spec}
\end{figure}

{The first general observable for SMBHs is their energy spectrum over large energy bands and it is therefore the first place to look for potential differences between 
a BBH and a single BH.}
{We first compute the full energy spectrum coming from the circumbinary disc as it} indicates to us in which frequency band those systems could be observed. {Contrary to the previous plots, those spectra are} 
computed over the full domain of the simulation {in order to be a good representative of the energy band even though the inner region is not our main focus here and therefore is not fully resolved}. 
  
Figure~\ref{fig:spec} shows the multi-color body emission spectrum of the circumbinary disc around an equal-mass BBH\footnote{These have been computed from spinning BBHs (with the corresponding {\tt GR-AMRVAC} simulations), as our metric construction allows us to. More details in \cite{mignon-risse_gravitational_2022}.}.
This figure illustrates how the peak of the energy spectrum is shifted to lower energies as the BBH mass increases \refe{(\citealt{rees_black_1984},} \citealt{alston_super-massive_2022}), while the overall flux increases.
Let us note that the emission model is defined aside from the self-consistent simulations.
Therefore, it allows for additional degrees of freedom.
While we have chosen to directly link the circumbinary disc maximal temperature to observational constraints \citep{lin_38_2013}, an alternative point of view is to derive the temperature from the cooling function used in the simulation \refe{(no such cooling function is used in the present work)}, as done in \cite{dascoli_electromagnetic_2018}.
Despite these distinct methods, we find an energy peak of the spectrum in rough agreement with the values obtained in \cite{dascoli_electromagnetic_2018} for a $10^{6}\Msol$ BBH (which would be located somewhere between our $10^5$ and $10^7\Msol$ cases).
Our peak is, however, located a much lower energies than the value of ${\sim}1$~keV${\sim}10^{17}$~Hz obtained analytically in \cite{tang_late_2018} in post-processing numerical simulations.
According to \cite{dascoli_electromagnetic_2018}, this discrepancy can arise from an accretion rate assumed to be ${\gtrsim}10^4$ greater in \cite{tang_late_2018} than theirs.
\\

\subsection{Timing features associated with the spiral}
\label{sec:lc}

\begin{figure}
	\includegraphics[width=\columnwidth]{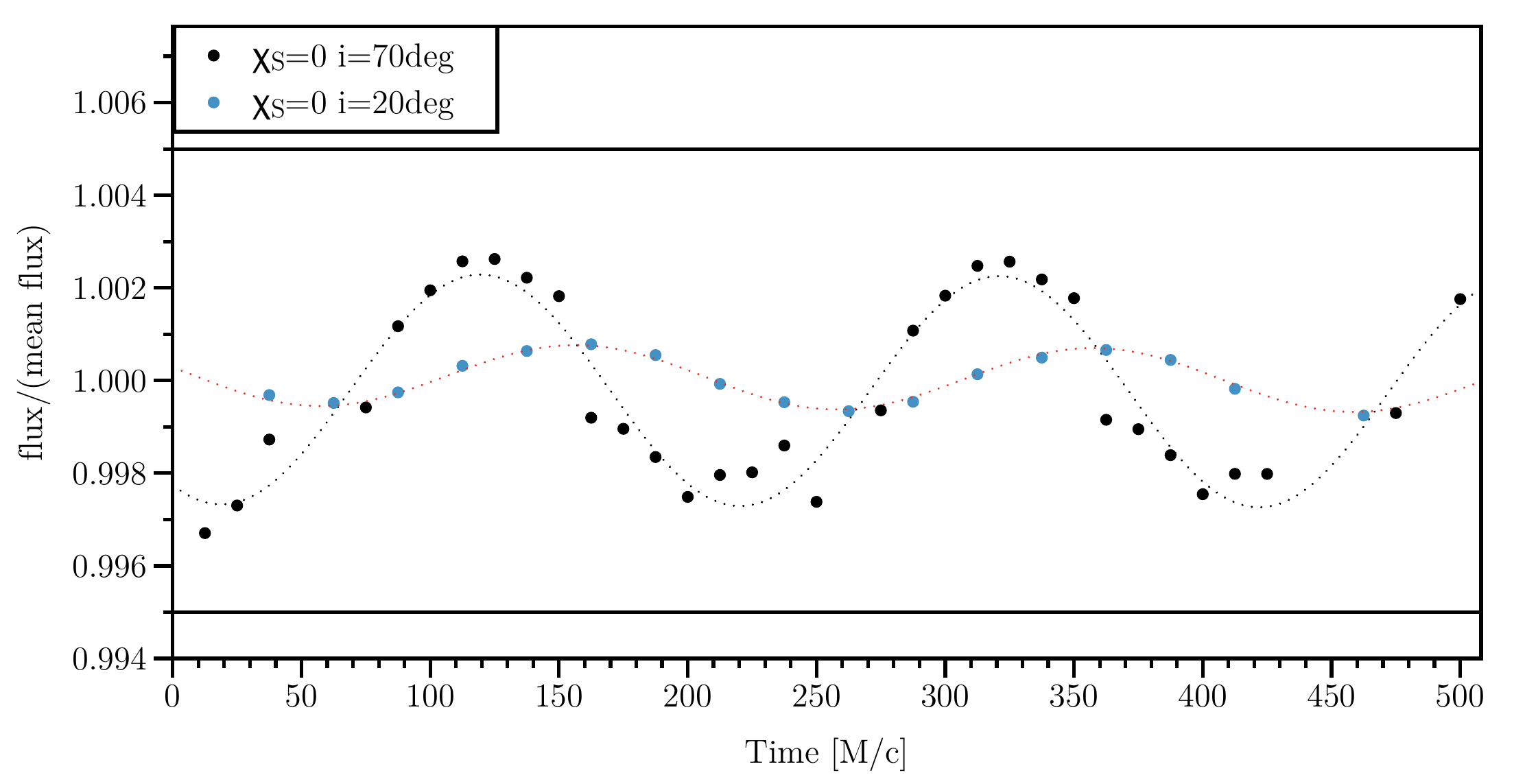}
    \caption{Flux divided by the mean flux, as a function of time in the $q=1$ case for {a system seen at} the inclination angles $20\deg$ and $70\deg$, where $0\deg$ means face-on.
    The orbital period is ${\approx}200$~M.}
    \label{fig:LCincli}
\end{figure}

While the direct impact of spiral waves on the energy spectrum is rarely \refe{detected in observations as it can be absorbed through other fitting parameters associated with the disc model} (\citealt{varniere_possible_2016}, \citealt{varniere_qpos_2022})
it is much easier to look at it in the Fourier space or, in the case of numerical simulation, by looking at the normalized lightcurve over a few periods.
{It is even better to look at high-inclination systems, as} spiral structures in discs produce timing features that depend {strongly} on the inclination \citep{varniere_impact_2016}.
Hence, we computed the lightcurve associated with the previous simulations to extract any observable timing feature having the potential to distinguish a BBH from a single BH.
When computing the lightcurves, we do not vary the total mass.
Indeed, as we renormalize the following lightcurves by the mean flux, any timing feature reported here is independent of the distance and of the total mass of the binary (naturally, the total flux is however determinant for detectability issues).
{\tt GYOTO} delivers a multi-frequency emission map of the source.
Integrating the emission map over the frequencies and over the source extent, we obtained each point of the lightcurve.

The lightcurve is displayed in Figure~\ref{fig:LCincli} in the equal-mass case for {the two extremes of low- and high-}inclination {systems}.
{As expected from the presence of the spiral}, we found that the lightcurve exhibits a clear modulation at the semi-orbital period of the binary (${\approx}200$~M).
We looked at the $20\deg$ inclination case as {a potential} worst case scenario to observe the impact of a spiral structure.
Moreover, the amplitude of the modulation increases with the inclination of the source. 
This trend is expected and is due to the modulation being produced by the {existence of a} non-axisymmetric density structure {in the disc}.
The amplitude of the modulation is rather weak, less than $1\%$, for inclinations $20\degree$ and $70\degree$.
\refe{This naturally depends on the energy band chosen; we chose an energy band optimal for the outer disc emission. Otherwise, the contamination from the tail of the inner disc emission is found to reduce this amplitude by roughly a factor of two.}

While this amplitude is small, it nearly reaches the order of magnitude of the weakest high-frequency quasi-periodic oscillations reported {in microquasars}  
{and techniques to search for such small quasi-periodic signals exist \citep{remillard_evidence_2002}, though applying them to low-flux sources such as the source of interest here is more difficult.

\subsection{Can we draw the inner mass distribution from the EM observables?}

\begin{figure}
	\includegraphics[width=\columnwidth]{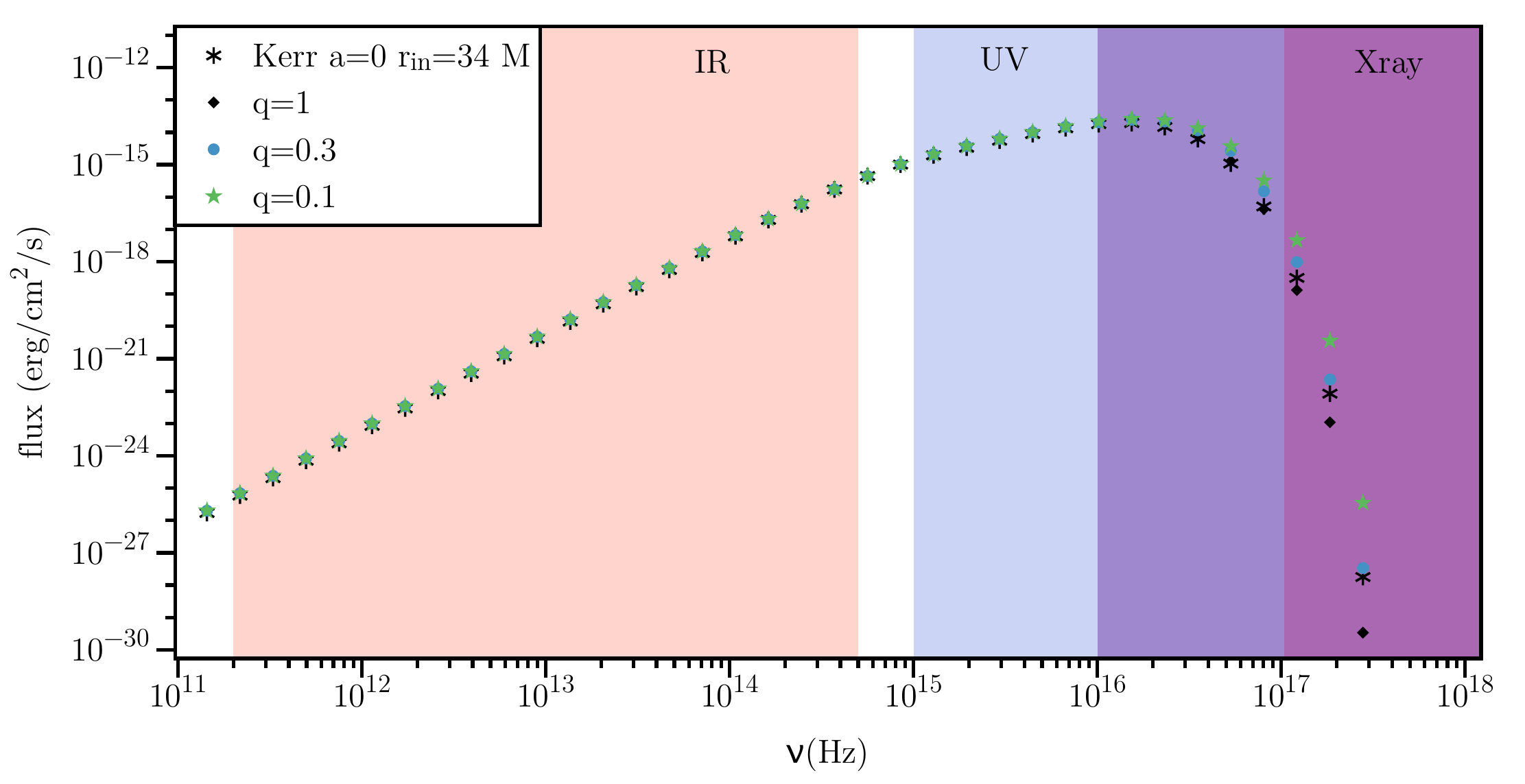}
    \caption{Energy spectrum at $t_\mathrm{8M}$, for $10^{5}\Msol$\refe{, for a single non-spinning BH and for BBHs} ranging mass ratios $q=0.1$ to $q=1$\refe{, at a distance of $500$~Mpc}.}
    \label{fig:spec2}
\end{figure}

{The previous plots were all made for the higher-amplitude case of equal-mass ratio and while it gives rise to a relatively small signal, it is not beyond what current timing capabilities have observed in low-mass systems.
It is therefore interesting to see how those features depend on the mass ratio and if they could be used to rule out some mass distribution inside the circumbinary disc,}
{especially focusing on the two extreme cases that go} from a single BH to an equal-mass BBH.

{We first focus solely on the energy spectrum which has the least impact from the spiral.}
Fig.~\ref{fig:spec2} {shows} the energy spectrum for various $q$ values and {the equivalent Schwarzschild} BH, with a total mass of $10^5\Msol$.
{As an extreme representation of the mass ratio, we choose to compare with a single} Schwarzschild BH {as the rest of the mass ratios were also done with no spin for the BBH}. 
{We also choose} an axisymmetric disc located at $34\mathrm{M}$, {to be coherent with} the radius of the circumbinary disc inner edge in the $q=1$ case\refe{; a disc blackbody spectrum is computed out of it, using the same temperature profile as before}.
{We see on Fig.~\ref{fig:spec2} that, while the spectrum in the X-ray shows some differences, all of them are compatible with the single-Schwarzschild case (since the flux differences and fluxes are small), which mean that the three mass-ratio SEDs
could be fitted by an equivalent Schwarzschild BH with a disc away from its last stable orbit. While such a far away disc would raise some questions as to what is pushing it away, it would require to 
have a precise measurement of the mass inside to be certain that the inner edge of the fitted disc is not at the last stable orbit. 
\refe{This is a somehow similar result to the spectrum deficit already reported in several papers (e.g. \citealt{gultekin_observable_2012} \citealt{shi_how_2016}) because of the gap cleared by the BBH (e.g. \citealt{artymowicz_dynamics_1994},\citealt{macfadyen_eccentric_2008}, \citealt{shi_three-dimensional_2012}, \citealt{noble_circumbinary_2012},
\citealt{dorazio_accretion_2013}, \citealt{farris_binary_2014},
\citealt{mosta_gas_2019},
\citealt{duffell_circumbinary_2020},
\citealt{noble_mass-ratio_2021},
\citealt{siwek_preferential_2022}).}
Exploring this in more details is beyond the scope of this paper\refe{, more focused on the radiation zone}, but it would be interesting to look at cases where we know that the inner edge of the disc in SMBH is far away from its last stable orbit and see if 
a BBH would be able to give a more physical explanation for such a far away disc.}
\newline

\begin{figure}
	\includegraphics[width=\columnwidth]{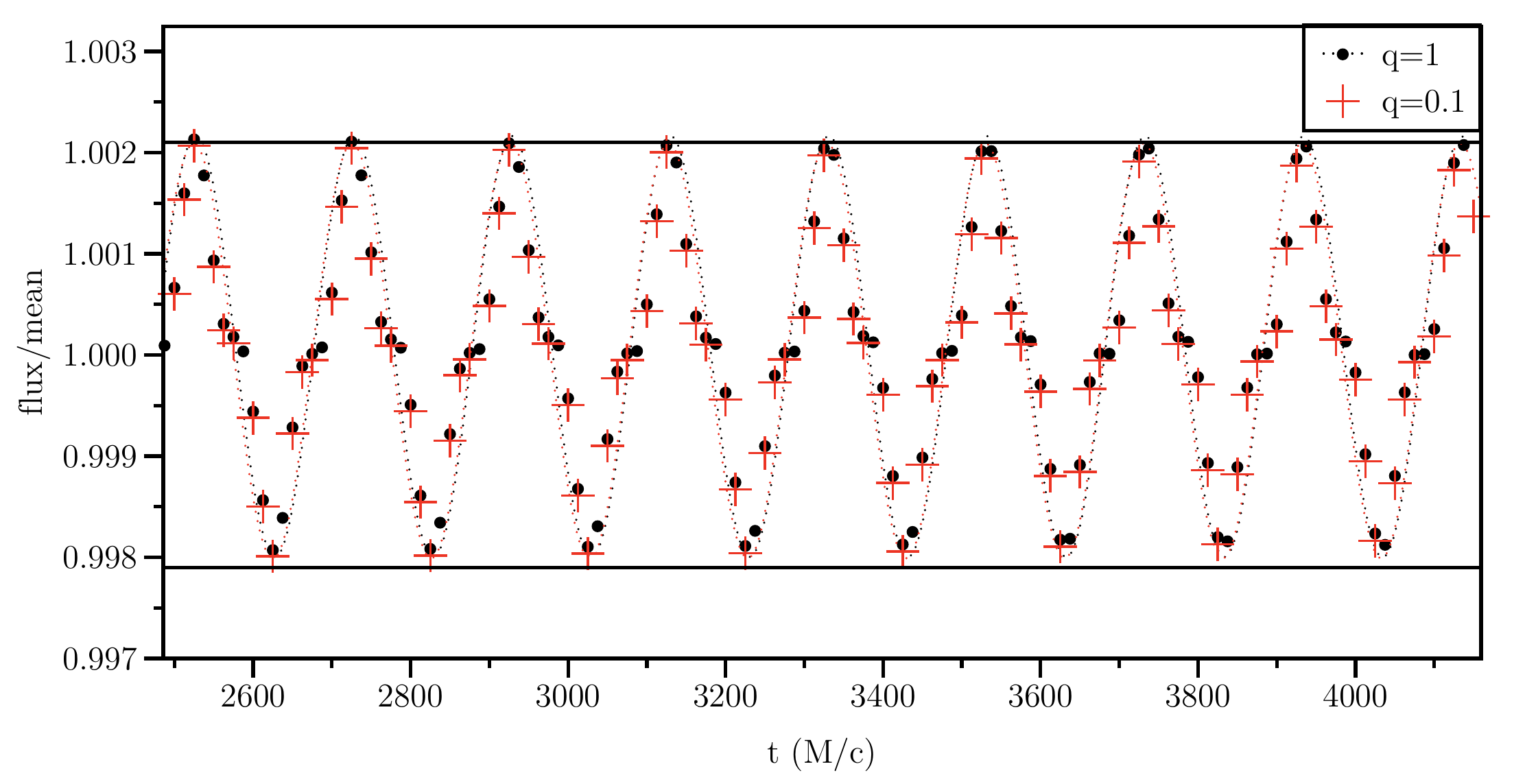}
    \caption{Flux divided by the mean flux, as a function of time, {for two mass ratios,} $q=1$ (black dots) and $q=0.1$ (red crosses).
    A sinusoidal curve has been added to the plot to show the periodic behaviour of the LC.
    The binary period is ${\approx}400$~M.}
    \label{fig:LCq}
\end{figure}

{As the spiral impact was more visible in the time domain, it is natural to} investigate how the mass ratio impacts the modulation seen in the lightcurve. {Hence, Figure~\ref{fig:LCq} shows the lightcurve for our three previous mass ratios in the case where the system is viewed at $60\deg$}.
Such an inclination angle is sufficient to observe the aforementioned modulation {while giving a more intermediate viewing angle.}
{As expected, the} period of the modulation is not sensitive to the mass ratio. 
{Indeed,} the mass ratio is not a leading-order effect in the theoretical orbital period (Eq.~4.2 of \citealt{johnson-mcdaniel_conformally_2009}).
{More surprisingly,}  the amplitude of the modulation is not affected by the mass ratio, 
{while we see it impacted} the temporal evolution of the amplitude of the spiral in the simulation (Fig.~\ref{fig:FNZrhot}).
It suggests that, due to the weak amplitude of the spiral structure in the disc (${\lesssim} 10^{-3}$), the major contribution to the modulation in the lightcurve comes from the existence of the spiral 
(i.e. a non-axisymmetry) \refe{and temporal variations along it,} rather than its exact amplitude. 
From theoretical modelling of the spiral impact in the disc, if the amplitude of the spiral structure were to reach a higher level ($\gtrsim1\%$), the impact {of the varying} amplitude on the lightcurve may have the potential to become detectable then {and could reach to differences between mass-ratio values}.

We showed the presence of a modulation in the lightcurve associated with the presence of the spiral structure produced by \refe{radiation zone effects}.
The properties of this modulation (amplitude, period) do not allow us to derive the $q$ parameter of the BBH.
Meanwhile, an axisymmetric disc around a non-GW emitting system, i.e. a single BH, would produce a {constant} lightcurve,  {hence} qualitatively different from those produced here.
Thus, while this modulation does not allow us to capture the {exact} mass ratio among BBHs, it could permit us to distinguish a BBH from a single BH. 

Overall, no spectral feature associated with the lightcurve modulation allows us to {directly} differentiate a BBH from a single BH {without having a good measure of the mass}.
Thus, in the outer regions of a circumbinary disc studied in this work, a timing study {has the best chance to} provide {direct} hints of a BBH, even though it does not allow us to {infer} the $q$ parameter.

\section{Conclusions}

This is the first paper of a series {of papers} in which we aim to study the pre-merger phase of BBHs {and, in particular, look for ways to characterize them}.
In order to perform this study and model a BBH spacetime, several numerical developments were done.
First, we extended {\tt GR-AMRVAC} \citep{casse_impact_2017}, a GR-MHD code, to compute the fluid dynamics {in any type of spacetime}.
Second, we implemented a BBH approximate analytical spacetime including the so-called Near and Far Zones (\citealt{johnson-mcdaniel_conformally_2009}, \citealt{mundim_approximate_2014}, \citealt{ireland_inspiralling_2016}) in {\tt GR-AMRVAC} and in the GR ray-tracing code {\tt GYOTO} \citep{vincent_gyoto_2011}.
Those two codes form the {\tt e-NOVAs} {framework we are using to} study  accreting compact objects, from fluid simulations to the production of synthetic observables, each step including {full} GR effects.
As the BBH spacetime relies on the trajectory of {each black hole}, we solved their EOM in the GW-dominated regime, as done in e.g., \citealt{combi_superposed_2021}.
Overall, we demonstrated the capability of {\tt GR-AMRVAC} and {\tt GYOTO} to handle non-stationary spacetimes.
Such developments are of major interest in the multi-messenger astronomy era to come.

In this first paper, we were interested in the potential impact of \refe{radiation zone effects, which combine the propagation of GWs emitted by inspiralling BBHs through their circumbinary disc and 
and the associated retardation effects.
First, we found a deviation in the lapse radial derivative with respect to the Schwarschild case, deviation that increases with distance and takes the form of a spiral because of the BBH trajectory.
This deviation cannot be reproduced with Newtonian retarded potentials.
We attribute it to the retardation effects in the BBH potential deriving from the diagonal part of its stress-energy tensor and contributing to the metric}.
We found that the \refe{radiation zone effects} create a steady spiral structure affecting the disc density, saturating at the $0.1\%$ level in the best case scenario, i.e. equal-mass BBH.
As expected from this small impact, we found that the energy spectrum is not sufficiently affected by the non-axisymmetric density distribution to leave a detectable imprint.
Nevertheless, the existence of a spiral creates a small - slightly smaller than $1\%$ - but potentially detectable modulation of the lightcurves at the semi-orbital period.
The amplitude of the modulation is not found to depend on the mass ratio (from $0.1$ to $1$), but,
as is generally the case for spirals, the modulation is larger at larger inclination angles (edge-on view of the disc).

\refe{In this series of papers, we aim} at predicting possible EM signals from inspiralling, GW-emitting, SMBBHs, to be detected with LISA.
The lightcurve modulation we report is precious in that view, as it is absent in axisymmetric disc models around single BHs.
While its amplitude is small, it nearly reaches the level of the weakest quasi-periodic oscillations detected (see \citealt{remillard_evidence_2002}).
Moreover, the timing features presented here are independent of the source mass and could apply to LIGO/Virgo/Kagra sources as well \citep{mignon-risse_gravitational_2022}.
By focusing on the outer parts of the circumbinary disc, our study is complementary to those investigating inner disc variabilities (e.g. \citealt{westernacher-schneider_multi-band_2021}, \citealt{gutierrez_electromagnetic_2022}) which will dominate in different energy bands.
\refe{Moreover, the effect we report is somehow more universal and permanent than features produced in the inner regions of the disc, such as spiral density waves.
Indeed, it is linked to how matter responds to the spacetime and to GWs propagating at the speed of light, while spiral density waves travel at the much slower local sound speed and are, consequently, also affected by the local thermodynamics.
In any case, the small amplitude oscillation we report can be understood as a lower-limit on the gravitational impact of a BBH on its circumbinary disc,\FC{ firstly because spiral density propagating outwards would increase the non-axisymmetry of the system and secondly because the amplitude of the effect we report  is likely to \refee{increase} as the binary gets closer to the merger.}}

\section*{Acknowledgements}

The authors thank the anonymous referee for his/her comments.
RMR thanks Fabrice Dodu for optimizing the computation of the metric components.
RMR also thanks Miguel Zilh\~ao and Hiroyuki Nakano for their insights on the equation of motion of inspiralling BBHs.
RMR thanks Sylvain Marsat for useful discussion.
RMR acknowledges funding from CNES through a postdoctoral fellowship.
This work was supported by CNES, focused on Athena, by the LabEx UnivEarthS, ANR-10-LABX-0023 and ANR-18-IDEX-000, by the "Action Incitative: Ondes gravitationnelles et objets compacts" and the Conseil Scientifique de l'Observatoire de Paris.
This work was granted access to the HPC resources of CINES under the allocation 2021-A0100412463 made by GENCI. 
Part of the numerical simulations we have presented in this paper were produced on the DANTE platform (APC, France).

\section*{Data Availability}

The data that support the findings of this study are available from the corresponding author, R.M.R, upon request.



\bibliographystyle{mnras}
\bibliography{Zotero} 




\appendix
\begin{table*}
\section{Frame transformation matrices}
\label{app:fido}
\justifying{In this appendix, we give the frame transformation matrices used to solve the Riemann problem in the locally Minkowski frame. 
Those matrices can be derived following the desired properties quoted in section 3.3 of \cite{white_extension_2016} (see also their Appendix A).
For usefulness, we give the matrices and inverse matrices in the three spatial directions and expressed in terms of the lapse $\alpha$, the shift components $\beta^i$ and the contravariant 3-metric components $\gamma^{ij}$, as implemented into {\tt GR-AMRVAC}.
In the first direction, the frame transformation matrix from the Minkowski basis to the coordinate basis is given by :
\begin{equation}
M^\mu_{\hat{\nu}} =
 \displaystyle\begin{pmatrix}
\frac{1}{\alpha} 		& 0 								& 0 				& 0 \\
-\frac{\beta^1}{\alpha} & \sqrt{\gamma^{11}} 					& 0 				& 0 \\
-\frac{\beta^2}{\alpha} & \frac{ \gamma^{12}}{\sqrt{\gamma^{11}}} 	& D \gamma_{33} 	& 0 \\
-\frac{\beta^3 }{\alpha} & \frac{ \gamma^{13}}{\sqrt{\gamma^{11}}}	 & -D  \gamma_{23} 	& \frac{1}{\sqrt{\gamma_{33}}}
\end{pmatrix}
\end{equation}

where $D = (\gamma_{33} (\gamma_{22}\gamma_{33}-\gamma_{23}\gamma_{23}) )^{-1/2}$. 
The corresponding inverse matrix is:
\begin{equation}
M^{\hat{\mu}}_\nu = \begin{pmatrix}
\alpha 																						& 0 											& 0 					& 0 \\
\frac{\beta^1}{\sqrt{\gamma^{11}}} 																	&  		\frac{1}{\sqrt{\gamma^{11}}} 				& 0 					& 0 \\
\frac{1}{D \gamma_{33}} \left[ \beta^2 - \frac{\beta^1 \gamma^{12}}{ \gamma^{11} } \right]  						& - \gamma^{12}/(\gamma^{11} D \gamma_{33}) 	& 1/(D \gamma_{33}) 	& 0 \\
\beta^3 \sqrt{\gamma_{33}	}+ \beta^2 \frac{\gamma_{23}}{\sqrt{\gamma_{33}}} - \frac{\beta^1}{\gamma^{11}} \left[ \gamma^{13}\sqrt{\gamma_{33}} + \frac{ \gamma^{12} \gamma_{23}}{\sqrt{\gamma_{33}}} \right]	& 
\frac{-1}{\gamma^{11}} \left[ \gamma^{13} \sqrt{\gamma_{33}} + \frac{\gamma^{12} \gamma_{23}}{ \sqrt{\gamma_{33}} }  \right]	 & \frac{ \gamma_{23} }{\sqrt{\gamma_{33}}} 	& \sqrt{\gamma_{33}}
\end{pmatrix}
\end{equation}

In the second direction, the frame transformation matrix from the Minkowski basis to the coordinate basis is given by :
\begin{equation}
M^\mu_{\hat{\nu}} =
 \begin{pmatrix}
\frac{1}{\alpha} 		& 0 						& 0 									& 0 \\
-\frac{\beta^1}{\alpha} & 1/\sqrt{\gamma_{11}}	&  \frac{ \gamma^{21}}{\sqrt{\gamma^{22}}} 	& - D \gamma_{31}  \\
-\frac{\beta^2}{\alpha} & 0 						& \sqrt{\gamma^{22}}					& 0 \\
-\frac{\beta^3 }{\alpha} &0	 					&  \frac{ \gamma^{23}}{\sqrt{\gamma^{22}}} 	&  D  \gamma_{11}  
\end{pmatrix}
\end{equation}

with $D = (\gamma_{11} (\gamma_{33}\gamma_{11}-\gamma_{31}\gamma_{31}) )^{-1/2}$.
The corresponding inverse matrix is:

\begin{equation}
M^{\hat{\mu}}_\nu = \begin{pmatrix}
\alpha 																& 0 										& 0 					& 0 \\

\beta^1 \sqrt{\gamma_{11}	}+ \beta^3 \frac{\gamma_{31}}{\sqrt{\gamma_{11}}} - \frac{\beta^2}{\gamma^{22}} \left[ \gamma^{21}\sqrt{\gamma_{11}} + \frac{ \gamma^{23} \gamma_{31}}{\sqrt{\gamma_{11}}} \right] 	
&  	 \sqrt{\gamma_{11}}				&  \frac{-1}{\gamma^{22}} \left[ \gamma^{21} \sqrt{\gamma_{11}} + \frac{\gamma^{23} \gamma_{31}}{ \sqrt{\gamma_{11}} }  \right] 					& \gamma_{31}/\sqrt{\gamma_{11}}  \\

\frac{\beta^2}{\sqrt{\gamma^{22}}}  											& 0	& \frac{1}{\sqrt{\gamma^{22}}}  & 0 \\

\frac{1}{D \gamma_{11}} \left[ \beta^3 - \frac{\beta^2 \gamma^{23}}{ \gamma^{22} } \right]  & 0	 & - \gamma^{23} /( \gamma^{22} D \gamma_{11}) 	&  1/(D \gamma_{11})
\end{pmatrix}
\end{equation}

In the third direction, the frame transformation matrix from the Minkowski basis to the coordinate basis is given by :
\begin{equation}
\centering
M^\mu_{\hat{\nu}} =
 \begin{pmatrix}
\frac{1}{\alpha} 		& 0 				& 0 						& 0 \\
-\frac{\beta^1}{\alpha} &  D  \gamma_{22}	& 0 						&  \frac{ \gamma^{31}}{\sqrt{\gamma^{33}}} \\
-\frac{\beta^2}{\alpha} & - D \gamma_{12} 	&  1/\sqrt{\gamma_{22}} 	& \frac{ \gamma^{32}}{\sqrt{\gamma^{33}}}  	 \\
-\frac{\beta^3 }{\alpha} & 0	 			&  0						& \sqrt{\gamma^{33}}
\end{pmatrix}
\end{equation}

with $D = (\gamma_{22} (\gamma_{11}\gamma_{22}-\gamma_{12}\gamma_{12}) )^{-1/2}$. 
The corresponding inverse matrix is:

\begin{equation}
M^{\hat{\mu}}_\nu = \begin{pmatrix}
\alpha 																& 0 							& 0 					& 0 \\

\frac{1}{D \gamma_{22}} \left[ \beta^1 - \frac{\beta^3 \gamma^{31}}{ \gamma^{33} } \right]  &  	1/(D \gamma_{22})			& 0					& - \gamma^{31} /( \gamma^{33} D \gamma_{22})   \\

\beta^2 \sqrt{\gamma_{22}	}+ \beta^1 \frac{\gamma_{12}}{\sqrt{\gamma_{22}}} - \frac{\beta^3}{\gamma^{33}} \left[ \gamma^{32}\sqrt{\gamma_{22}} + \frac{ \gamma^{31} \gamma_{12}}{\sqrt{\gamma_{22}}} \right]  																					&  \gamma_{12}/\sqrt{\gamma_{22}}	&  \sqrt{\gamma_{22}}	&  \frac{-1}{\gamma^{33}} \left[ \gamma^{32} \sqrt{\gamma_{22}} + \frac{\gamma^{31} \gamma_{12}}{ \sqrt{\gamma_{22}} }  \right]  \\

\frac{\beta^3}{\sqrt{\gamma^{33}}} 											& 0	 						& 0					&  \frac{1}{\sqrt{\gamma^{33}}} 
\end{pmatrix}
\end{equation}
}\end{table*}


\bsp	
\label{lastpage}
\end{document}